\documentclass[prb,showpacs,superscriptaddress,twocolumn,aps,10pt]{revtex4-1}
\usepackage[utf8]{inputenc}
\usepackage[american,british]{babel}
\usepackage[T1]{fontenc}
\usepackage[pdftex]{graphicx}  
\usepackage{graphicx, xcolor}
\usepackage{dcolumn}
\usepackage{bm}
\usepackage{amsmath,amsthm,amssymb}
\usepackage{multirow}
\usepackage{tabularx} 
\usepackage{booktabs}
\usepackage{colortbl}
\usepackage{amsmath}
\usepackage{color}
\usepackage{verbatim}
\usepackage{ulem}

\definecolor{darkGreen}{RGB}{0,110,0}
\definecolor{darkBlue}{RGB}{0,0,130}
\usepackage[colorlinks,citecolor=darkGreen,linkcolor=darkBlue,urlcolor=blue,hyperindex]{hyperref}

\definecolor{white}{rgb}{1.0, 1.0, 1.0}
\definecolor{ghostwhite}{rgb}{0.97, 0.97, 1.0}

\definecolor{teagreen}{rgb}{0.82, 0.94, 0.75}
\definecolor{turquoisegreen}{rgb}{0.63, 0.84, 0.71}
\definecolor{mossgreen}{rgb}{0.68, 0.87, 0.68}
\definecolor{cottoncandy}{rgb}{1.0, 0.74, 0.85}
\definecolor{lavenderpink}{rgb}{0.98, 0.68, 0.82}
\definecolor{palegreen}{rgb}{0.6, 0.98, 0.6}
\definecolor{pastelpink}{rgb}{1.0, 0.82, 0.86}
\definecolor{pink}{rgb}{1.0, 0.75, 0.8}
\definecolor{babypink}{rgb}{0.96, 0.76, 0.76}
\definecolor{grannysmithapple}{rgb}{0.66, 0.89, 0.63}
\newcommand\colourpadding[1]{\addlinespace[-1pt]\arrayrulecolor{#1}\midrule[1pt]\arrayrulecolor{black}}

\usepackage{float}

\usepackage{caption}
\usepackage{subcaption}

\begin{document}

\title{Identifying vegetation patterns for a qualitative assessment of land degradation using a cellular automata model and satellite imagery}
\author{Hediye Yarahmadi}
\email{hediye.yarahmadi@tcd.ie}
\affiliation{School of Physics, Trinity College Dublin, Dublin 2, Ireland}
\author{Yves Desille}
\email{yves.desille@universite-paris-saclay.fr}
\affiliation{Université Paris-Saclay, 91405 Orsay, France}
\affiliation{School of Physics, Trinity College Dublin, Dublin 2, Ireland}
\author{John Goold}
\email{gooldj@tcd.ie}
\affiliation{School of Physics, Trinity College Dublin, Dublin 2, Ireland}
\author{Francesca Pietracaprina}
\email{francesca.pietracaprina@tcd.ie}
\affiliation{School of Physics, Trinity College Dublin, Dublin 2, Ireland}

\begin{abstract}

We aim to identify the spatial distribution of vegetation and its growth dynamics with the purpose of obtaining a qualitative assessment of vegetation characteristics tied to its condition, productivity and health, and to land degradation. To do so, we compare a statistical model of vegetation growth and land surface imagery derived vegetation indices. Specifically, we analyze a stochastic cellular automata model and data obtained from satellite images, namely using the Normalized Difference Vegetation Index (NDVI) and the Leaf Area Index (LAI). In the experimental data, we look for areas where vegetation is broken into small patches and qualitatively compare it to the percolating, fragmented, and degraded states that appear in the cellular automata model.
We model the periodic effect of seasons, finding numerical evidence of a periodic fragmentation and recovery phenomenology if the model parameters are sufficiently close to the model's percolation transition. We qualitatively recognize these effects in real-world vegetation images and consider them a signal of increased environmental stress and vulnerability.
Finally, we show an estimation of the environmental stress in land images by considering both the vegetation density and its clusterization.
\end{abstract}

\maketitle

\section{Introduction}

Land degradation is a complex environmental process involving the loss of biological activity and economic productivity. It is caused both by peculiarities of the land and climate, and by un-adapted human activity.
Specifically, desertification is a transition that involves land degradation of drylands, resulting in the transformation of productive land into arid areas \cite{vonhardenberg2001diversity, shnerb2003reactive, kefi2007local, kefi2007spatial}.
Drylands\cite{myfootnote}, covering approximately $41\%$ of the Earth's land surface, are characterized by limited and fluctuating rainfall, exerting an environmental pressure, or stress, on soil and vegetation growth \cite{hobbs1992disturbance}. Several areas in the world, including Europe - which is the geographical focus of this work -, have seen an increase in dryland extent~\cite{spinoni2015}.

Monitoring large land areas is fundamental to understand the progression of these environmental changes. To this aim, numerous studies \cite{reynolds2007global, sala2000global, archer2001trees}  have explored the relationships among drylands, vegetation patterns, and external stresses, using a number of models and data sources. These include remote sensing \cite{jensen2009remote}, which uses satellite or airborne sensors to gather data on vegetation cover and health; geographic information systems  \cite{goodchild1991geographic} for analyzing spatial patterns and changes in vegetation; ecohydrological modeling \cite{tague2004rhessys} to examine the interaction between vegetation and water resources; agent-based modeling  \cite{grimm2006standard} to simulate individual-level behavior and its impact on ecosystems; and species distribution modeling  \cite{elith2006novel} to predict plant species distribution under different environmental conditions.
A statistical physics approach, alongside these methods, also contributes to understanding vegetation distribution and dynamics, as well as the role of disorder in the vegetation processes. Particularly, vegetation patchiness is a valuable tool for assessing degradation risk and detecting early warning signals associated with it~\cite{levick2013, dahlin2013, asner2013forest}. Disorder enters the vegetation dynamics through environmental fluctuations, stochastic processes such as weather patterns, disturbances (e.g., wildfires, human activities), and spatial heterogeneity in soil composition or topography. The presence of disorder affects both the existence and the way the transitions occurs, making them more gradual or continuous \cite{villa2015eluding, martin2014quenched, vazquez2011temporal}. This, in turn, may affect the environmental resilience. 

Our objective in this work is to describe the structure of vegetation cover through a numerical model of vegetation growth dynamics, and to conduct an analysis of satellite image data to recognize patterns of vegetation patchiness that are signs of land degradation in progress. We use a Stochastic Cellular Automaton (SCA) model~\cite{kefi2007spatial, kefi2007local,kefi2010bistability,kefi2014early}, which incorporates events such as vegetation mortality, survival and propagation.
This SCA model is born out of individual plant lifecycle considerations (e.g. propagation of seeds, new plant colonization, plant competition, etc). These effects are coarse-grained into stochastic state transitions of macroscopic cells which are arranged in a grid, covering large land areas. The state changes in the cells are able to generate patterns observed in arid ecosystems, such as gaps, stripes, and labyrinth-like structures \cite{kefi2007spatial, kefi2007local}. It also presents two phase transitions as the environmental stress (or vegetation mortality) changes: a desertification transition where vegetation disappears, and a percolation transition where the vegetation breaks down into clusters and which acts as a precursor to the desertification transition~\cite{corrado2014early, manor2008facilitation, von2010periodic}.

A number of data sources are available to survey and identify changes in vegetation and land ecosystems. These include satellite images from several Landsat missions, Sentinel, MODIS, SPOT and RapidEye~\cite{ose2016}, as well as databases of specific-purpose processed data (e.g. several vegetation intensity indices in the Copernicus Land~\cite{copernicus} database). The availability of such data has enabled the investigation of phase transitions in drylands~\cite{meloni2019vegetation}. Here, we aim to verify whether the processes in the SCA - namely the clusterization processes acting under increased environmental stress - can be qualitatively detected in satellite image data at a large scale, focusing on several areas across Europe.

The work is structured as follows: Sec. II provides an overview of the theoretical and numerical methods employed, including the model used for simulations and the incorporation of seasonal effects. In Sec. III, the analysis is applied to recent satellite images of a selection of European countries. Section IV connects the SCA simulated process and the observed land data. Finally, our conclusions are outlined in Sec. V.

\section{Vegetation dynamics and clusterization}
\label{sec:simulations}

In order to capture the vegetation patchiness dynamics, we consider the already mentioned SCA model introduced in ref.~\cite{kefi2007spatial} and used to study the dynamics of vegetation patterns in ecological systems.
In the following, we will provide a brief overview of the model, while we refer to the original papers for a comprehensive description \cite{kefi2007spatial, kefi2010bistability}.

We represent the vegetation by means of a three-state SCA model. In a $L \times L$ square lattice, each cell can exist in one of three states: (+) is a vegetation-covered state (living cell); (0) is an empty state, available for colonization (dead cell); and (-) is a degraded state (degraded cell).
Cells undergo state transitions with transition rates $W_{ij}$. For instance, $W_{0+}$ represents the rate of transition from an empty to a vegetated state. Not all transitions have a nonzero probability. For instance, a degraded cell must undergo recovery before it can be colonized. At the same time, only an empty cell is vulnerable to degradation. Consequently, transitions between a degraded state (-) and a vegetated state (+) are prohibited. The rates at which the allowed transitions take place are as follows:
\begin{align}
\label{eq:mortality}
W_{+0} &= m \\
\label{eq:colonization}
W_{0+} &= [\delta \rho_+ + (1 - \delta)q_{+|0}](b - c\rho_+) \\
\label{eq:degradation}
W_{0-} &= d \\
\label{eq:recovery}
W_{-0} &= r + fq_{+|-}
\end{align}
The system's dynamic evolution is governed by a Markov chain, with equations \eqref{eq:mortality} to \eqref{eq:recovery} representing mortality, colonization, degradation, and recovery processes, respectively. Transitions from and to the dead (0) state are influenced by $q_{i|j}$: given a cell in state $j$, this represents the fraction of its nearest neighbors in state ${i}$. Besides the contribution of the neighboring cells, the equations (\ref{eq:mortality})-(\ref{eq:recovery}) involve additional parameters related to the lifecycle of plants, with the following interpretations: $\delta$ is the proportion of seeds dispersed by wind, animals, etc.; $b$ is the colonization parameter, which accounts for various intrinsic properties of a vegetated cell, such as seed production rate, seed survival, germination, and survival probabilities (not including global competition effects); $c$ is the strength of global competition effects; $d$ represents the rate of soil degradation, incorporating intrinsic soil characteristics, climatic factors, and anthropogenic influences; $f$ is the local facilitation parameter, describing cooperative interactions among plants and positive feedback between soil and plants; and finally, $r$ is the spontaneous regenerative rate of a degraded cell in the absence of vegetation covering the neighboring cells. The parameter values used in the following analysis are: $b = 0.6$, $c = 0.3$, $d = 0.2$, $\delta = 0.1$, $f = 0.9$, and $r = 0.0004$. These values reflect typical processes in semiarid ecosystems (particularly: intermediate colonization, low competition, intermediate soil degradation, high facilitation, and low spontaneous regeneration) and align with those used in previous studies in order to reflect real field data simulations \cite{corrado2014early, kefi2007spatial, kefi2014early}.
A wide range of parameter values has been analyzed in the literature with numerical simulations in order to qualitatively reproduce different climatic regimes~\cite{kefi2007local}.
In our analysis, the control parameter for the model is the mortality rate $m$, which represents the intensity of external stress.

The process to simulate vegetation dynamics is as follows: we start by initializing a randomized lattice configuration consisting of alive, barren, and dead cells. The system evolves through the transition probabilities given by equations (\ref{eq:mortality})-(\ref{eq:recovery}). We note that, while eq. \eqref{eq:mortality} and \eqref{eq:recovery} can be applied independently, eq. \eqref{eq:colonization} and \eqref{eq:degradation} have to satisfy the condition $W_{00}+W_{0+}+W_{0-}=1$, on whether the zero cells will transition into the alive or barren states or preserve their current situation.
After a transient dynamics due to the initialization of the system with a random configuration, the system converges to an equilibrium which depends on the chosen parameter set. This initial transient dynamic is discarded (typically $5000$ iterations); once the system reaches equilibrium, we collect the vegetation fraction and vegetation cluster sizes for each iteration step.
Throughout this work, unless explicitly specified, the simulations are conducted on square $L \times L$ lattices with a linear size of $L = 100$ with periodic boundary conditions. This system size is comparable with the sample size chosen for the satellite images in Section~\ref{sec:sec3} and has weak finite size effects~\cite{corrado2014early}. The time series data typically consists of at least $10^{4}$ records, but near the desertification threshold, the number of steps is increased up to $10^{6}$ due to critical slowing down effects.

The properties of vegetation clusters are investigated through the alive vegetation density $\rho_+=\frac{N_+}{L^2}$, where $N_+$ is the number of alive cells, and the size of the largest cluster of alive cells $\mathcal{C}_+$, which is the largest vegetation cluster size divided by the total number of lattice cells. We analyze a percolation transition in response to the external stress parameter $m$.

\subsection{Numerical simulations of vegetation dynamics}

\begin{figure}
	\centering
	\includegraphics[width=\columnwidth]{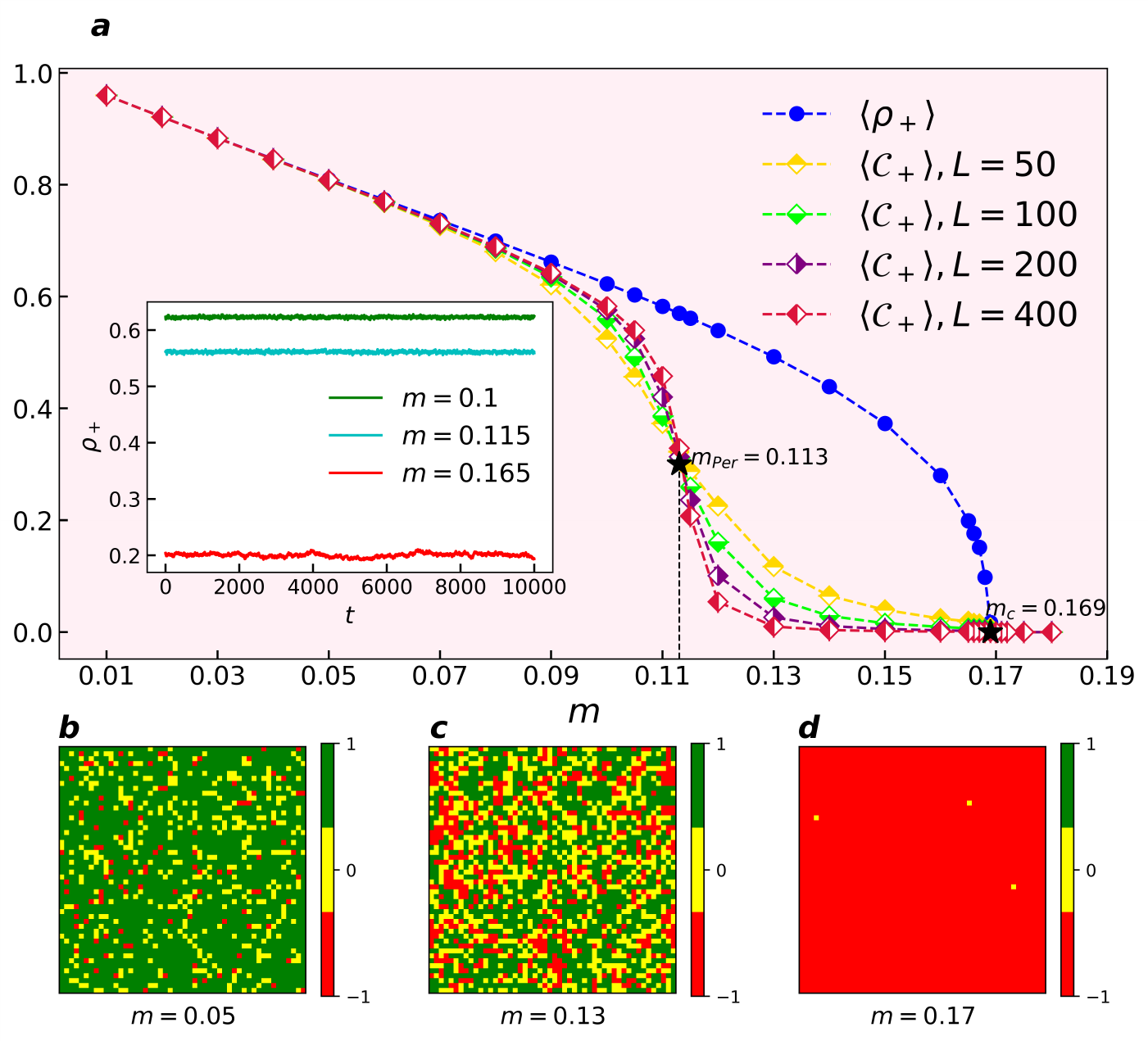}
	\caption{{Evolution of vegetation dynamics.} Panel \textbf{(a)}: The average of the vegetation density $\langle \rho_+ \rangle$ for $L=100$ and the relative size of the largest cluster $\langle \mathcal{C}_+\rangle$ for $L=50, 100, 200, 400$ as a function of the mortality rate $m$. The markers represent the percolation threshold $m_{Per}=0.113$, which is determined from the crossing point for the various system sizes, and the degradation transition point $m_c=0.169$. Inset of \textbf{(a)}: The typical evolution of the living cell density $\rho_+(t)$ for several values of the mortality rate $m$. The linear lattice size is $L = 100$, and the time unit corresponds to one iteration step. Panels \textbf{(b)-(d)}: Vegetation distribution in a lattice of size L = 50 at \textbf{(b)} low (m = 0.05), \textbf{(c)} intermediate (m = 0.13), and \textbf{(d)} high (m = 0.17) mortality. Green cells are vegetated, yellow cells are empty, and red cells are degraded areas.}
	\label{Fig1}
\end{figure}

In this section, we investigate the model for different values of the mortality rates $m$. 
The inset in Fig.~\ref{Fig1}\textbf{(a)} shows the evolution of the living cell density $\rho_+(t)$, for several values of $m$. After a short transient (not shown), the system fluctuates around an equilibrium value $\rho_+(m)$. With the increase of external stress $m$, the average density of vegetated cells decreases, until a continuous phase transition occurs at $m_c = 0.169$ (see the main panel of Fig.~\ref{Fig1}\textbf{(a)}) corresponding to a desertification transition. Additionally, we determine the size of the largest cluster of vegetated cells $\mathcal{C}_+(m)$. As $m$ increases, at finite sizes $\mathcal{C}_+$ has a sharp crossover that corresponds to a percolation transition which acts as a precursor to desertification~\cite{corrado2014early, kefi2007spatial, kefi2014early}. Considering different system sizes allows to find the percolation transition point at $m_{Per} = 0.113$ as the crossing point of the curves at different $L$. See Sec. I of Supplemental Material for information about the transitions.

Thus, two distinct transitions are involved in land degradation processes~\cite{corrado2014early,corrado2015desertification,corrado2015critical,bengochea2022habitat}. The first transition, a percolation transition in the living vegetation clusters, serves as an early warning sign for the second transition, which is associated with desertification as the fraction of living vegetation goes to zero. 
Both of these transitions have been observed to occur in ecosystems~\cite{von2010periodic}. The bottom panels of Fig.~\ref{Fig1} shows snapshots of the model at different stages: healthy and percolating (panel \textbf{b}), vulnerable and broken into small clusters (panel \textbf{c}), and degraded (panel \textbf{d}).

\subsection{Modeling the seasonal effects}

So far, we have considered a SCA model with a constant mortality rate. In order to include the effect of seasonal cycles in real ecosystems, we introduce a time-dependent, periodic mortality rate $m(t)$. 
We aim to model a simplified time dependence in the environmental stress. To do so, we consider an asymmetric periodic function representing a fast increase in environmental stress, which is reminiscent of the phenomenological temperature and precipitation time series in semiarid areas\cite{footnote2}.

Specifically, we are interested in investigating whether an ecosystem that seasonally is put in a situation of high environmental stress is able to recover, as well as what conditions need to be satisfied to avoid a permanent transition to a desertified state.

In our analysis, starting from a randomly generated configuration, we first let the system equilibrate with a constant mortality parameter $m_\text{in}=m_0$. At time $t_\text{in}$ we apply a time-dependent $m(t)$ given by the following asymmetric functional form, which we choose as it is both simple in terms of number of parameters, and resembles phenomenologically the experimental time series of air temperature in the geographic areas considered:

\begin{equation}
        m(t) = 
            m_0 + A\sin[2\pi/T \,t + k \sin(2\pi/T \,t)] 
            \hspace{0.5cm}
            t > t_\text{in}
\label{eq:time-dependent-mortality}
\end{equation}

$m(t)$ consists of a constant component, $m_0$, representing the average environmental stress across the seasons. The additional parameters in the oscillatory term are: the amplitude $A$, representing the intensity of the seasonal effect, the period $T$ representing the duration of a year in the simulation's arbitrary time units, and an asymmetry parameter $k$ representing fast season changes particularly from winter to summer. Here we consider $T = 5000$ and $k=0.2$.
Note that $T$ is sufficiently large to be greater than the time scale of the internal fluctuations of $\rho_+$. The asymmetry $k$ is chosen to resemble fits of experimental temperature data. The initial value $m_0$ is set to several values near the percolation and degradation transition points, $m_\textit{Per}=0113$ and $m_c=0.169$ respectively.
Values just below $m_c$ are specifically interesting in order to determine the effects of periodic increases in environmental stress and whether they can drive a transition to a non-recoverable dead or barren state. 
The numerical computation is performed for $m_0=0.08$ to $0.17$, varying $A$, and $10^6$ time steps.

\begin{figure}
	\centering
    \includegraphics[width=\columnwidth]{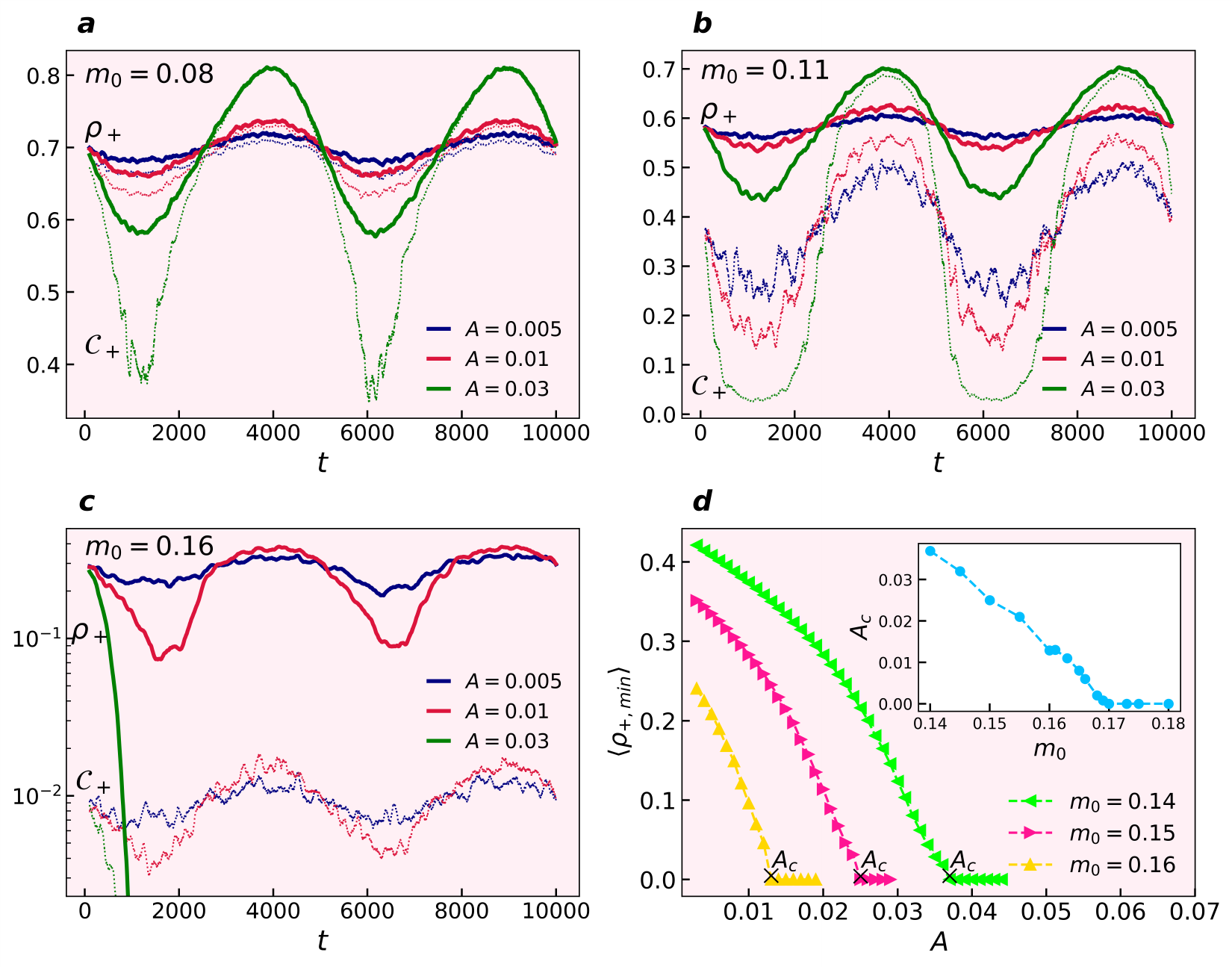}
    \caption{{Evolution of vegetation dynamics for time-dependent mortality rate $m(t)$.} Panels \textbf{(a)}, \textbf{(b)} and \textbf{(c)}: Vegetation density $\rho_+$ (solid lines) and largest cluster size $\mathcal{C}_+$ (dotted lines) as a function of time, for initial mortality $m_0=0.08, \,0.11$ and $0.16$ respectively. Data is smoothed by means of a moving average with a window size of $100$.
    Panel \textbf{(d)}: Mean value of the minima of the vegetation density $\langle \rho_{+,min} \rangle$ as a function of the amplitude $A$. We mark the crossover amplitudes $A_c(m_0, T)$ where the curves drop to $0$.
    Inset of panel \textbf{(d)}: Crossover amplitude $A_c$ as a function of $m_0$.}
	\label{Fig2}
\end{figure}

We investigate the impact of increasing the amplitude $A$ on the behavior of $\rho_+$ and $\mathcal{C}_+$.
Fig.~\ref{Fig2}\textbf{(a)} shows the effects of different amplitude values for $m_0=0.08$. $\rho_+$ shows a periodical response to $m(t)$, with a very small response delay in the system size considered. 
$\mathcal{C}_+$ closely follows and overlaps with $\rho_+$ for the majority of the time, showing that most alive cells belong to the same contiguous cluster. For sufficiently high amplitudes, $\mathcal{C}_+$ temporarily dips to lower values only near the minimums of $m(t)$, with a fast recovery to values overlapping $\rho_+$.

In Fig.~\ref{Fig2}(\textbf{b}) we show results for a higher mortality which is close to the percolation transition value and periodically sent above it. In this case as well, $\mathcal{C}_+$ and $\rho_+$ both show a periodical response to $m(t)$. For low amplitudes, $\mathcal{C}_+$ is at all times significantly lower than $\rho_+$ while still being non-zero. For sufficiently high amplitude, however, it quickly goes to very small values during the lower half-period of the oscillation, suggesting that the system has been temporarily driven to the highly fragmented and non-percolating regime; although $\mathcal{C}_+$ stays near 0 for a substantial part of the lower half-period, a similar rapid recover happens in the upper half-period, up to reaching values close to $\rho_+$ only near the maximum of $m(t)$.

Finally, in Fig.~\ref{Fig2}\textbf{(c)} we show the results for a system oscillating around $m_0=0.16$, that is a point close to the transition value $m_c$.
Interestingly, we show that even when the time-dependent mortality rate goes above the critical value, the system can recover its alive vegetation cover $\rho_+$ and continue to respond to the oscillating $m(t)$, as long as the amplitude of the oscillation, and thus the time spent in a condition where $m(t)>m_0$, is sufficiently small. We find that there is a crossover amplitude $A_c$ such that for $A>A_c$ the system enters a state of almost all dead or degraded cells and is no longer able to recover, even though the environmental stress $m(t)$ strongly decreases afterwards.
The drop to a dead or degraded configuration happens as the system fluctuates near the crossover point with increasingly long metastable oscillating states. At this value of $m_0$, the system is already firmly in the highly fragmented regime and $\mathcal{C}_+$, while responding to the oscillations alongside $\rho_+$ (as long as the system is not fully degraded and $\rho_+$ is not identically $0$), is an order of magnitude smaller.

Next, we extract the crossover amplitude $A_c$ for several values of average mortality $m_0$. To do so, we look for the time series which identically reach $0$.
For each value of $A$, we determine the value of $\rho_{+,min}$, the minimum vegetation density for each period, which happens at the time of highest seasonal stress for each period. We use a cubic spline interpolation to determine the minima for each period. $\rho_{+,min}(t)$ reaching $0$ quantitatively marks the drop to a non-recoverable degraded configuration.
Figure~\ref{Fig2}\textbf{(d)} shows $\langle \rho_{+,min} \rangle$, the average minimum across all periods, as a function of the amplitude $A$ for $m_0=0.14, 0.15$, and $0.16$. The crossover to full degradation is observed as $\langle \rho_{+,min}(A) \rangle$ goes to zero at $A=A_c$. For $A>A_c$, $\rho_{+}$ will be asymptotically zero after an unstable transient. Finally, in the inset of Fig.~\ref{Fig2}\textbf{(d)}, we plot these crossover amplitude points $A_c(m_0)$.
$A_c$ is approximately linear in $m_0$ until $m_0=0.17\approx m_c$, at which point it reaches zero. Beyond this point, for $m_0\geq 0.17$, $A_c$ is identically zero.

The system's behavior and crossover to desertification are influenced by the period $T$, with shorter oscillations (compared to the system's own intrinsic time scales) resulting in a more robust alive phase.

\section{Finding vegetation clusters through satellite data}
\label{sec:sec3}

Data obtained from earth observation satellites can be analyzed to recognize the patterns of vegetation cover in real ecosystems. In this section, we recognize areas of fragmented vegetation qualitatively matching the phases exhibited by the cellular automaton model of Section~\ref{sec:simulations}.

We use two vegetation indices commonly used to detect vegetation from satellite imagery: the Normalized Difference Vegetation Index (NDVI) \cite{pettorelli2013normalized} and the Leaf Area Index (LAI) \cite{fang2019overview}. NDVI is widely used in ecosystem monitoring due to its simple formulation. It measures the vegetation greenness by calculating the ratio of spectral reflectances in near-infrared and red light. Green, living plants have a high reflection in the near-infrared and high absorption in red and visible light frequencies, which is a markedly different response than bare soil, water, snow, or urbanized areas. For this reason, it is commonly used in remote sensing to qualitatively assess vegetation health and density in a specific area.
We also use the LAI data, which is defined as half the total area of living vegetation elements in the canopy per unit of ground area. This quantity is obtained either through direct measurement (performed locally on a sampling basis) or through indirect methods such as image analysis. All of the data for our analysis is extracted from the European database \textit{Copernicus Global Land Service}, which provides global vegetation data products~\cite{copernicus}.

The NDVI and LAI images were acquired from the Copernicus Land dataset for three days per month: for NDVI on the $1^{st}$, $11^{th}$, and $21^{st}$ day of each month spanning the time period from $2014$ to $2021$, while for LAI on the $10^{th}$, $20^{th}$, and last day of each month, from $2014$ up until August $2020$ (there are gaps in the LAI data during the winter and autumn seasons, due to insufficient illumination in satellite observations). Both datasets have a resolution of 300m and the data used is composed of a mosaic corrected for atmospheric differences, including the removal of cloud coverage. Our analysis focuses on specific regions, namely in France, Germany, Ireland, Spain, and Greece. We note that drylands are particularly present in Spain and Greece.
We pre-processed the images in the dataset in order to normalize the individual image pixel data, a greyscale image with values from 0 to 255, to $[0,1]$ for land areas. Additionally, we exclude pixels corresponding to sea, lakes, rivers and any water bodies that have been pre-marked by the Copernicus dataset preprocessing, as well as those otherwise detectable through NDVI and LAI intensity values: indeed, some smaller water areas need to be manually excluded, which we do through a local thresholding method (we use the Threshold-Sauvola approach \cite{sauvola1997adaptive}, see the
SM Sec. II for more detail). We note that the values of $\mathcal{C}_+$ and $\rho_+$ which will be computed from the experimental data and used in the following analysis have been normalized to the land area only, excluding water bodies.

\subsection{Spatial distribution of vegetation and breakdown into clusters}
\label{sec:sec3_breakdown}

\begin{figure}
	\centering
	\includegraphics[width=\columnwidth]{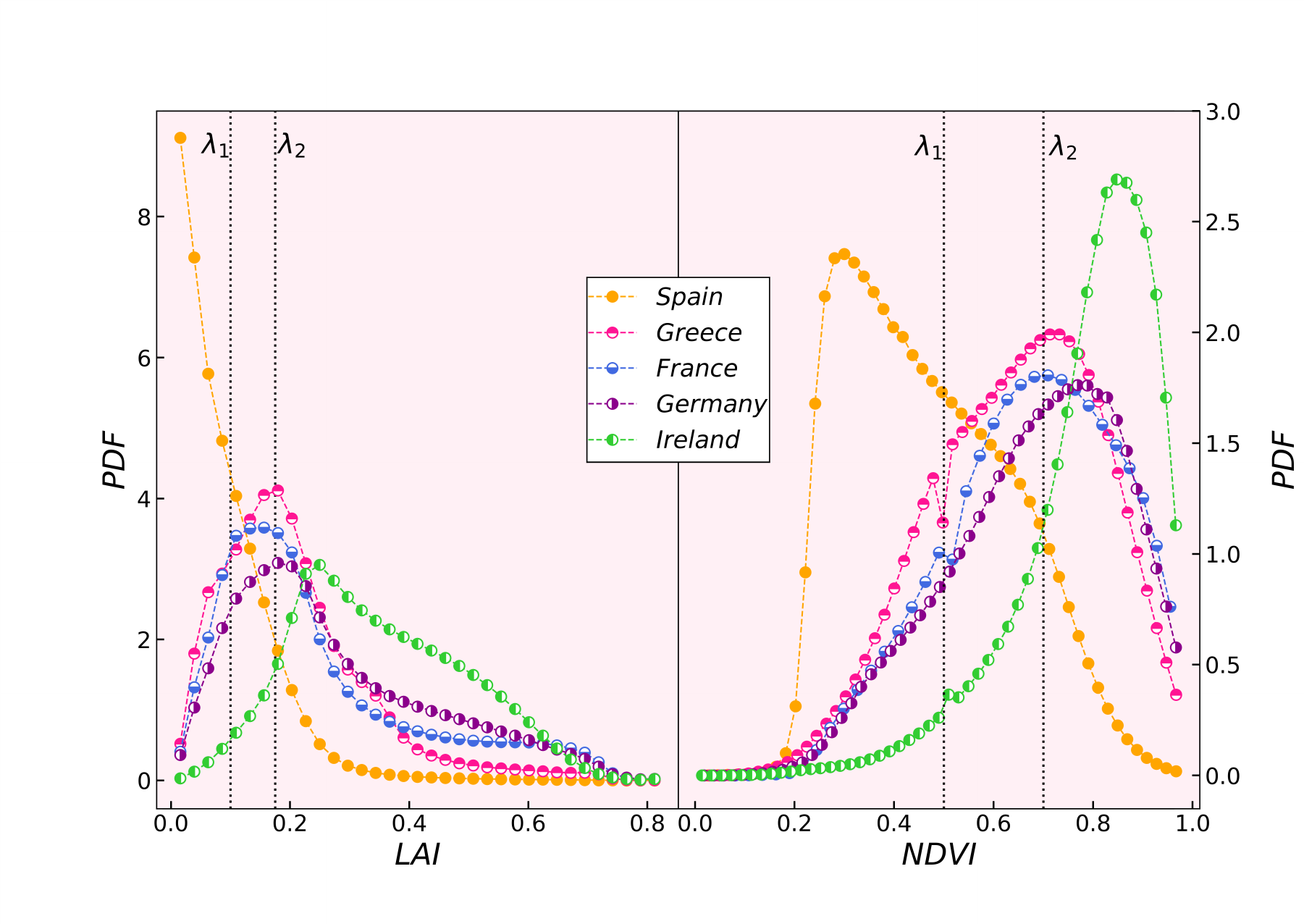}
	\caption{Probability distribution of normalized NDVI and LAI values across examined land regions. The data is aggregated over samples taken at different dates and times. As a reference, the threshold values $\lambda_1$ and $\lambda_2$ for Greece, France, and Germany are marked as vertical dotted lines.}
	\label{Fig3}
\end{figure}

We now analyze the satellite images of vegetation in order to identify a breakdown into disconnected clusters. Specifically, here we look for the breakdown in connectivity which in the vegetation model used in Section~\ref{sec:simulations} translates to a percolation transition.
To do so, we match the pixel intensity in the satellite image to three possible states, reminiscent of the SCA model. We thus introduce a simplification to discretize the data, which for a single pixel corresponds to an intensity between 0 and 1, based on two chosen threshold parameters. This effectively maps low intensity pixels to a degraded cell, medium intensity to an empty cell, and high intensity to an alive cell. We are then able to consider the fraction of alive vegetation $\rho_+$ and the largest vegetation cluster $\mathcal{C}_+$ analogously to the analysis of the SCA model in Section~\ref{sec:simulations}.

Determining the threshold values is tied to the typical values of the NDVI and LAI indices, which itself are qualitative in nature, and their interpretations in terms of plant health.
As a starting point, we consider the Probability Distribution Function (PDF) of NDVI and LAI pixels intensity across the analyzed regions, shown in Fig.~\ref{Fig3}. It is worth noting that LAI is a curated index, deriving from a mix of indirect and direct measurements, and it is not directly and quantitatively comparable to NDVI. Both indices clearly signal the presence of vegetation, with different distributions due to the climate and morphological characteristics of the country groups: especially intense values in Ireland, and less vegetation intensity in Spain.

We introduce the two cut-off parameters $\lambda_1$ and $\lambda_2$, which classify the pixels as vegetated/alive for values $>\lambda_2$, empty/dead for values between $\lambda_1$ and $\lambda_2$ and degraded for values $<\lambda_1$.
The specific values of $\lambda_1$ and $\lambda_2$ are chosen so that they are respectively below and around the typical values of the indices. Note that for different regions we may choose different $\lambda_1$ and $\lambda_2$ values due to differences in vegetation type and thus NDVI or LAI intensity; this is particularly relevant for the selected areas in Spain and Ireland. In the following analysis, we use the values $\lambda_1=0.5$, $\lambda_2=0.7$ (NDVI) and $\lambda_1=0.1$, $\lambda_2=0.175$ (LAI) for Greece, France and Germany, all of which have similar NDVI and LAI intensities.
We note that the qualitative results of the following analysis will not depend on the precise values chosen (see SM Sec.~V for more details).

The SCA model considered in Section~\ref{sec:simulations} describes lifecycle processes at the scale of a single plant; it is however also suitable to coarse-graining, as the mortality, reproduction/colonization, degradation and recovery processes can be applied collectively and averaged over land areas. Our aim is to verify that the overall behaviors of in the simplified SCA model can be qualitatively detected in vegetation even at the rather coarse resolution of 300m of the considered satellite data.

The NDVI and LAI images include, as well as water bodies which are identified and excluded from each sample as described above, areas which are blocked from vegetation growth, and therefore locked in a ``degraded'' state, due to anthropic causes (e.g. cities). As well, additional human activities such as farming and grazing will introduce systematic effects in the classification of the image pixels into the three states. In the following analysis, we neglect these effects on the analysis of vegetation clusterization.
Moreover, we neglect the internal heterogeneity of the data within each region, namely the diversity in soil type, local climate, prevalent plant type and any other small-scale properties. This choice is due to both the low resolution of the source image data and of the selected samples (which, as outlined in Section~\ref{sec:sec3_analysis}, will have a linear size of 30km), and the simplified coarse-grained nature of the analysis of vegetation dynamics that will be performed.

\subsection{Analysis of NDVI time series}
\label{sec:sec3_analysis}

\begin{figure}
	\centering
	\includegraphics[width=\columnwidth]{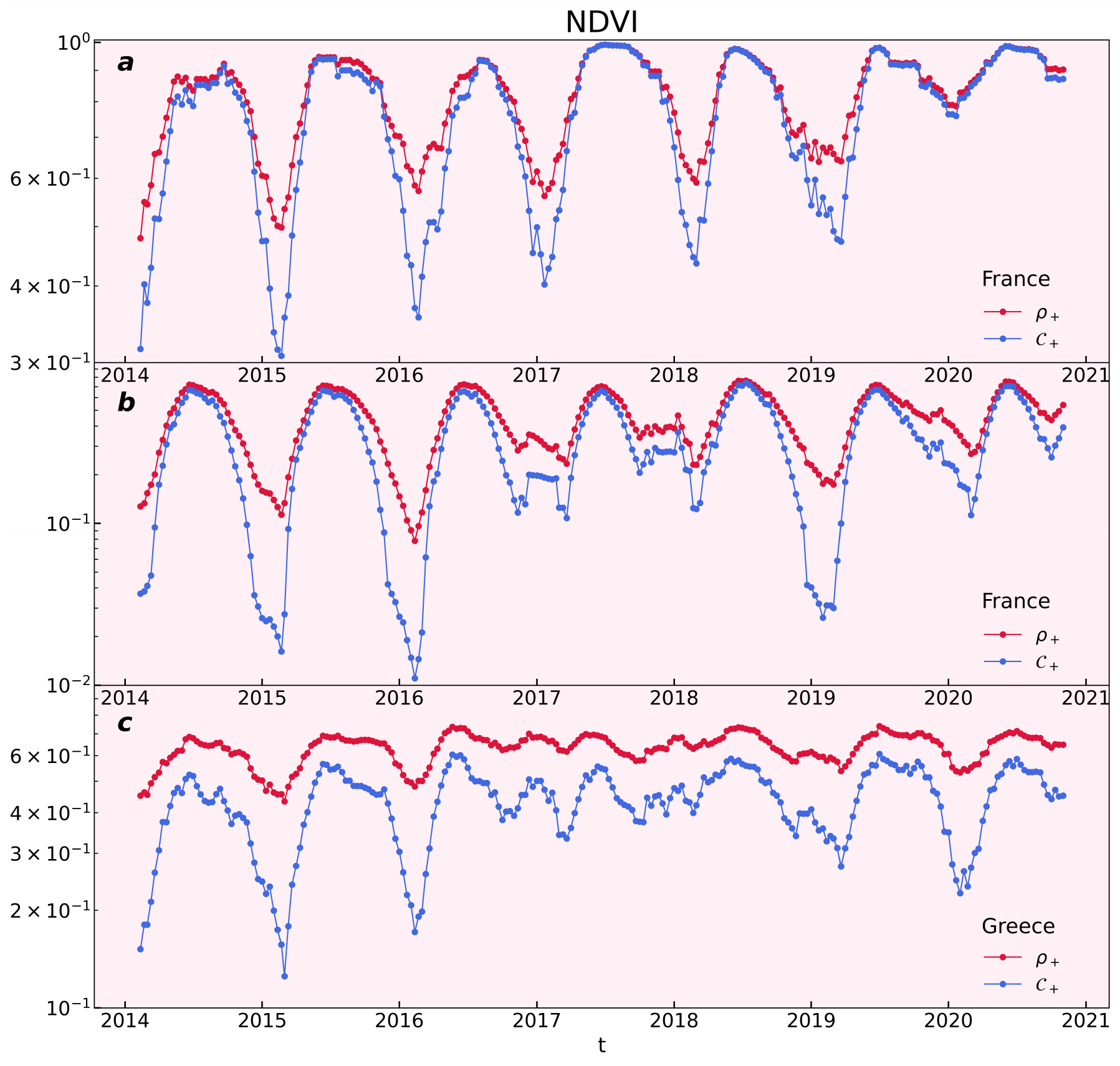}
	\caption{Comparison of $\rho_+$ and $\mathcal{C}_+$ as a function of time for three samples, (\textbf{a}) and (\textbf{b}) in France and (\textbf{c}) in Greece. We use NDVI data for the years 2014 to 2021. The data is smoothed by moving average over a window of size 10.
		}
	\label{Fig4}
\end{figure}

Here we focus on NDVI data specifically for areas in France (first two samples) and Greece (last sample), which respectively have a higher or lower prevalence of semiarid conditions. The areas that we consider in each different country is split in samples of $100\times100$ pixels (that is, 30km$\times$30km). For each sample, we analyze the fraction of alive vegetation $\rho_+$ and the relative size of the largest vegetation cluster $\mathcal{C}_+$ over time. 
Among the samples considered, we see a range of phenomenology, which we analyze qualitatively and for which we show some examples in Fig.~\ref{Fig4}.

In Fig~\ref{Fig4}(\textbf{a}) we consider a sample with very high maximum values of $\rho_+(t)$ and $\mathcal{C}_+(t)$ and where, for the majority of the time, there is an overlap between $\rho_+(t)$ and $\mathcal{C}_+(t)$. This indicates that nearly all the vegetated pixels belong to the largest cluster of vegetation. This situation is akin to the phase at a low mortality rate in the simulated vegetation dynamics in Section~\ref{sec:simulations}, which we associate to healthy vegetation and low environmental stress.
We note a seasonal periodic behavior in both $\rho_+(t)$ and $\mathcal{C}_+(t)$ with a temporary dip in $\mathcal{C}_+$ during the autumn or winter season. This is qualitatively similar to the behavior in Fig~\ref{Fig2}(\textbf{a}) for low average mortality rate in the presence of strong seasonal oscillations.

In Fig~\ref{Fig4}(\textbf{b}) we show a sample exhibiting a more marked difference between $\rho_+(t)$ and $\mathcal{C}_+(t)$, which only overlap at their maximum points. This situation is similar to Fig~\ref{Fig2}(\textbf{b}), where a mortality rate close to the percolation transition value is periodically pushed above it.

Finally, in Fig~\ref{Fig4}(\textbf{c}) we show a sample with consistently different $\rho_+(t)$ and $\mathcal{C}_+(t)$, which themselves exhibit low values. This is qualitatively comparable with Fig~\ref{Fig2}(\textbf{c}), where the average mortality rate is between the percolation and the degradation point and, if seasonal oscillation have a sufficiently small amplitude, $\mathcal{C}_+(t)$ oscillates as $\rho_+(t)$ but around a much smaller average value without pushing the system to full degradation.

We have repeated this analysis for the additional areas that have been considered in Spain, Germany, and Ireland. As for Greece and France respectively, we are able to identify a higher prevalence of situations of stress associated with the semiarid climate for Spain, and generally its absence in Germany and Ireland.

\section{Mapping vegetation stress}

In this section, we qualitatively compare the vegetation fraction and clusterization data obtained from the SCA model simulations with the experimental one obtained from satellite images.
We aim here to identify the areas that show vulnerability: these areas will have a relatively dense vegetation (high $\rho_+$), but are non-contiguous and broken down into many small clusters (low $\mathcal C_+$). As shown in section~\ref{sec:simulations}, this is a percolation transition that precedes the situation where vegetation is not surviving. Increased vulnerability is especially detectable in many areas during the autumn and winter months, as observed by the analysis of the time series of $\rho_+$ and $\mathcal{C}_+$ in Section~\ref{sec:sec3_analysis}.

In order to better visually identify the situation in which each image sample lies, we consider a simultaneous scatter plot of $\rho_+$ and $\mathcal{C}_+$. We first simulate the SCA model for several (non-time-dependent) values of the mortality rate $m$ and we construct the curve $(m, \overline{\rho_+}, \overline{\mathcal{C}_+})$. The averages $\overline{\rho_+}$ and $\overline{\mathcal{C}_+}$ are computed by taking the centroid of the scatter cloud obtained from a large ($10^4$ - $3\cdot 10^5$) number of snapshots of the equilibrated system. Interestingly, the projection of the curve in the $\rho_+, \mathcal{C}_+$ plane is independent of the value of the parameters of the SCA model for a wide range of realistic parameters (see
Sec. III of the SM).
Given their universality, we use this simulated curve as a reference for the $\rho_+$ and $\mathcal{C}_+$ computed from the experimental image data.

We then again consider the sub-image samples of size $30km\times 30km$ and consider the values of $\rho_+$ and $\mathcal{C}_+$ for each sample. We summarize again here the three main scenarios.
Firstly, when we encounter areas with high $\rho_+$ and $\mathcal{C}_+$, it is an indication of regions with low mortality rate: this corresponds to a large portion of the image covered by unbroken, contiguous percolating vegetation. The second case corresponds to regions characterized by low $\rho_+$ and $\mathcal{C}_+$ values. These areas thus have low vegetation coverage, which include genuinely degraded land, rocky terrains, and urban areas. The last case involves instances of high $\rho_+$ values but low $\mathcal{C}_+$ values. This combination suggests that despite the presence of relatively dense vegetation, the environmental stress and thus the mortality rate is high, causing the vegetation to break down into small clusters, as a precursor to full degradation.
The data points obtained from satellite images can be compared with two guidelines. The first is the already mentioned curve numerically simulated with the SCA model at various values of the mortality rate; the other is the line $\rho_+ = \mathcal{C}_+$, which corresponds to a fully connected vegetation. This condition represents a low mortality rate situation, and all data points will lie in the $\rho_+ > \mathcal{C}_+$ sector.

\begin{figure}
	\centering	\includegraphics[width=\columnwidth]{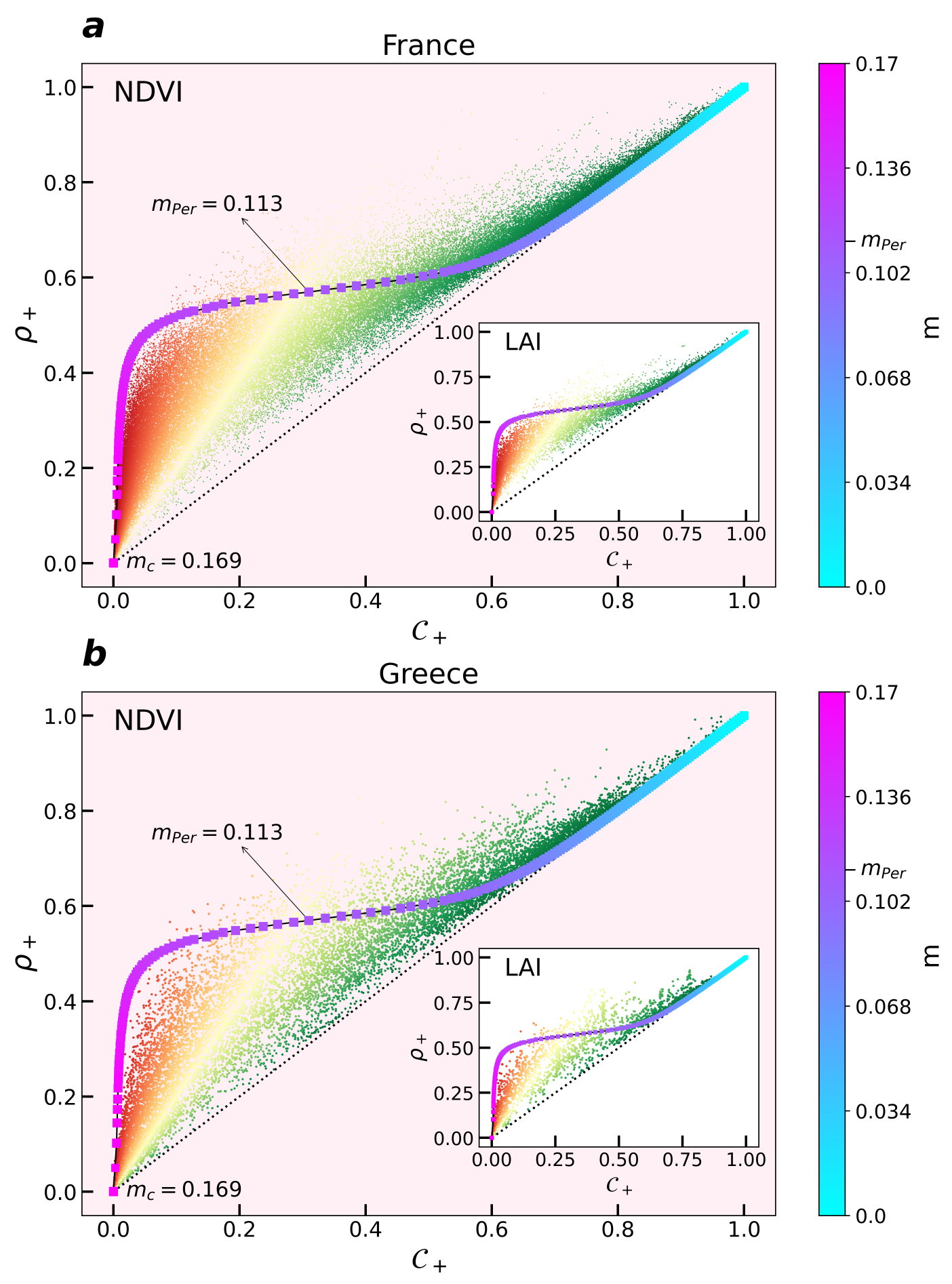}	\caption{Vegetation density $\rho_+$ and the relative size of the largest cluster $\mathcal{C}_+$ for $30km\times30km$ samples. The main panels \textbf{(a)} and \textbf{(b)} present the cumulative NDVI data for all available dates spanning the years 2014-2021, showing the results for France and Greece respectively. The points are color-coded with $\mathcal C_+/\rho_+$ representing a qualitative classification of the degradation in each sample. The diagonal $\rho_+ = \mathcal{C}_+$ is shown, as well as the $(m, \rho_+, \mathcal{C}_+)$ data points obtained from the numerical simulation of the SCA model with the parameters' values outlined in Section~\ref{sec:simulations}; the values of $m$ are represented through a color scale. Insets: results from the LAI data for all available dates spanning the years 2014 to 2021, for France (\textbf{a}) and Greece (\textbf{b}).
 }
	\label{Fig5}
\end{figure}

\begin{figure}
    \centering
    \begin{minipage}[b]{0.25\textwidth}
        \includegraphics[width=\textwidth]{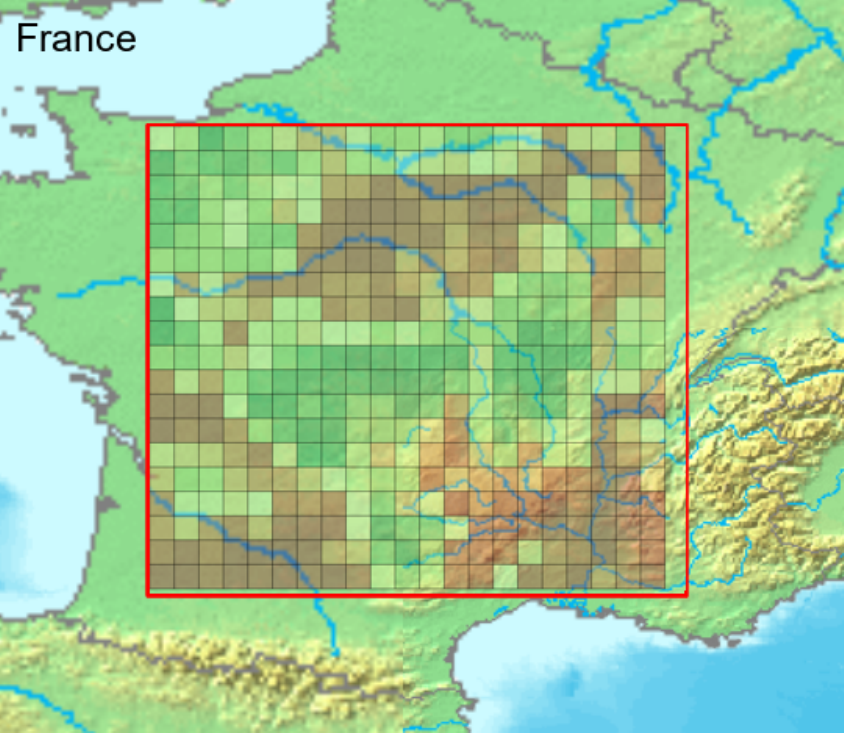}
    \end{minipage}
    \begin{minipage}[b]{0.208\textwidth}
        \includegraphics[width=\textwidth]{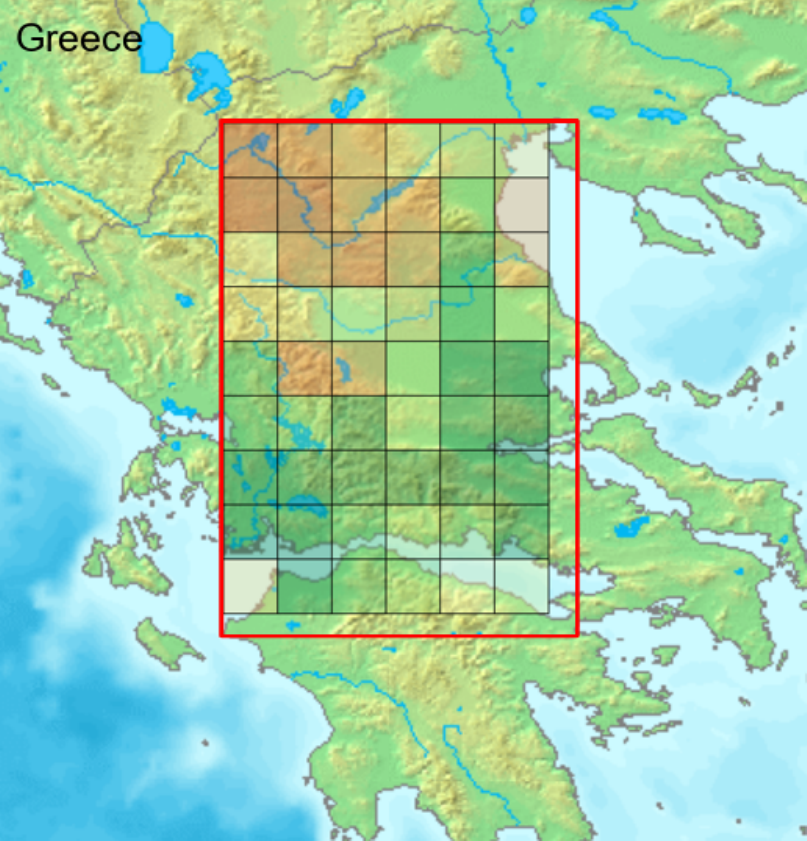}
    \end{minipage}
    \caption{Qualitative assessment of land degradation, obtained by comparing each sample's vegetation density and largest vegetation cluster size. The color coding refers to the value of $\mathcal C_+/\rho_+$ for each sample, representing a qualitative classification of its degradation. We show here France (left) and Greece (right) on 11 March 2020.}
    \label{Fig6}
\end{figure}

We show in Fig.~\ref{Fig5} a scatter plot of $\rho_+$ against $\mathcal{C}_+$ for all NDVI image samples obtained for France (panel \textbf{a}) and Greece (panel \textbf{b}) from 2014 to 2021; we additionally show the results for the LAI dataset in the inset. In this figure we display all available sample points, aggregated for all dates, in order to highlight their distribution in the $(\rho_+, \mathcal{C})$ plane.
For additional geographical areas, we refer to the SM Sec. IV.
The data points in Fig.~\ref{Fig5} approximately lie between the numerical curve obtained from the numerical simulations of the SCA model and the diagonal $\mathcal{C}_+ = \rho_+$. In order to understand the significance in terms of environmental stress, we should first highlight the systematic effect of permanently unvegetated pixels in each sample. These pixels might correspond to locations not covered by soil, e.g. urban or rocky areas. These reduce the effective available size of each sample, in a way that is not accounted by the model (as these pixels are not recoverable even in the most favorable conditions). This explains the presence of data points close to the diagonal for even mid and low values of $\mathcal{C}_+$. As well, points spread between the diagonal and the numerical curve correspond to increasingly degraded samples with reduced effective size. One could attempt to quantify this systematic effect by introducing stochastically arranged blocked pixels in the numerical model; we leave this detailed analysis to future work. Here, we give a simple qualitative assessment of vulnerability by evaluating the ratio $\mathcal C_+/\rho_+$.
Given its value, we classify the data points in three levels of vegetation health. The healthiest vegetation samples are located close to the diagonal and in the top right area (green points), where the vegetation is above the percolation transition. The points in the bottom left area (red points), close to the numerical curve, are samples whose vegetation is broken down into small non-percolating clusters, which therefore have the lowest vegetation health. 
The points in the central area have a reduced effective soil availability; among them, the points close to the diagonal (green points) still present contiguous vegetation, while, farther away from the diagonal and closer to $\mathcal C_+=0$, the samples show intermediate (yellow points) to low (red points) vegetation health. In Fig.~\ref{Fig5} we mark the different classifications for all points with green-yellow-red colors, for which we consider these approximate threshold values for $\mathcal C_+/\rho_+$: 0.3 for red to yellow and 0.7 for yellow to green.
Finally, we show this qualitative classification by color coding each sample with its $\mathcal C_+/\rho_+$ value in Fig.~\ref{Fig6} for areas in France and Greece, superimposed to a map of the land that has been analyzed, where each square correspond to one $30 \mathrm{km}^2$ sample, as defined in Sec.~\ref{sec:sec3_analysis}.

\section{Conclusions}

In this work, we have analyzed the spatial distribution of vegetation through a numerical stochastic model and through the analysis of land data from satellite images. The model is known to exhibit three phases, namely showing contiguous, fragmented, and degraded vegetation, and the corresponding two transitions, a percolation and a degradation transition which happen as the environmental stress increases. Considering samples of 30km in size, we looked at the vegetation fraction and the fraction of contiguous vegetation in each sample with the aim of identifying these processes in real land data.

Although the model has simplifications and the experimental data has relatively low resolution, we have shown that the model's description of the vegetation dynamics can be qualitatively compared to the land data. Particularly, we extended the model with a time-dependent environmental stress resembling the periodic seasonal changes. Some key aspects of its phenomenology can be recognized in the experimental time series for a variety of samples. Finally, the values of the vegetation fraction and the largest cluster size allowed us to offer a qualitative categorization of each sample between the three scenarios of contiguous, fragmented and degraded vegetation. As a future perspective, we highlight the potential usage of this classification as a tracker of land health (see an example of this use in the SM Sec. VI). Namely, after accounting for seasonal variations, the time series of vegetation fraction and cluster size for a land area can be used to detect worsening or recover of vegetation over time based on the historical data presented in this work.

\section*{Acknowledgments}
FP has received funding from the  European  Union’s  Horizon  2020  research and innovation program under the  Marie Sklodowska-Curie grant agreement No 838773.  JG is supported by a SFI-Royal Society University Research Fellowship and acknowledges funding from European Research Council Starting Grant ODYSSEY (Grant Agreement No. 758403). This work has received support from ERC PoC "Emerald", grant agreement No 101069222. Funded by the European Union. Views and opinions expressed are however those of the authors only and do not necessarily reflect those of the European Union or ERCEA. Neither the European Union nor the granting authority can be held responsible for them. The Irish Centre for High End Computing (ICHEC) has supported the computational aspects of this work through grants tcphy205c and tcphy223c.

\clearpage
\newpage

\section*{Supplementary Material}
\renewcommand\thefigure{S\arabic{figure}}   
\setcounter{figure}{0} 
\setcounter{section}{0}


\section{Additional information on the transitions in the SCA model}\label{Text-S1}

\begin{figure}[H]
    \centering
    \includegraphics[width=\columnwidth]{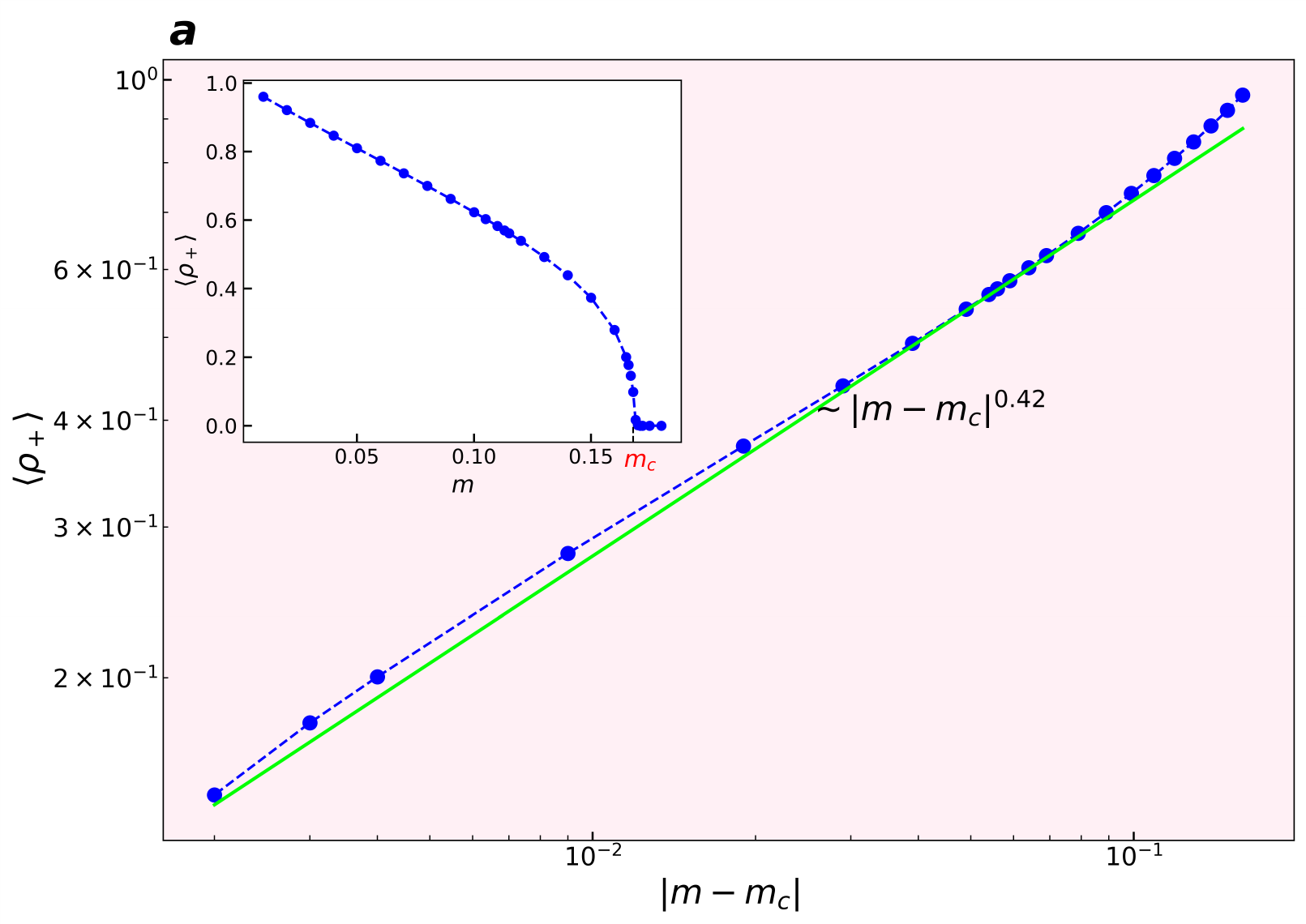}
    \hfill
    \includegraphics[width=\columnwidth]{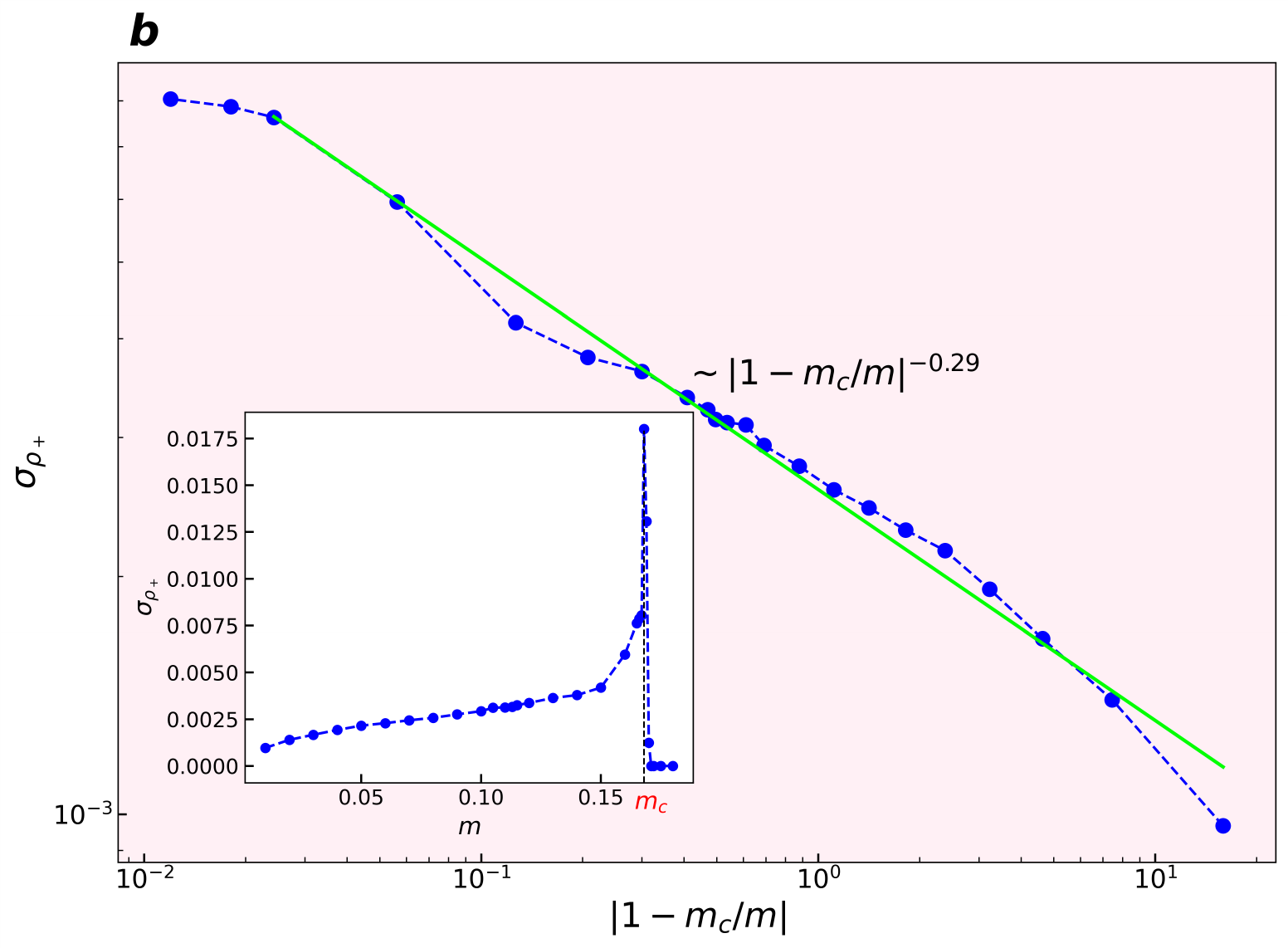}
    \caption{Panel \textbf{(a)} 
        Main: A plot of $\langle \rho_+ \rangle$ vs $|m - m_c|$ in log-log scale. The green solid line represents the best-fit line, corresponding to a power law with a slope of $\beta_1=0.42 \pm 0.01$. Inset: The average density of living cells $\langle \rho_+ \rangle$ as a function of $m$ on a linear scale.
        Panel \textbf{(b)} 
        Main: A plot of $\sigma_{\rho_+}$ vs $|1 - m_c/m|$ in log-log scale. The green solid line represents the best-fit line, corresponding to a scaling law with a slope of $-\gamma^{(1)}_{\sigma}=-0.29 \pm 0.01$. Inset: The root-mean-square deviation of the living cell density $\sigma_{\rho_+}$ as a function of mortality rate $m$ on a linear scale.}
    \label{figS1}
\end{figure}

\begin{figure}
    \centering
    \includegraphics[width=\columnwidth]{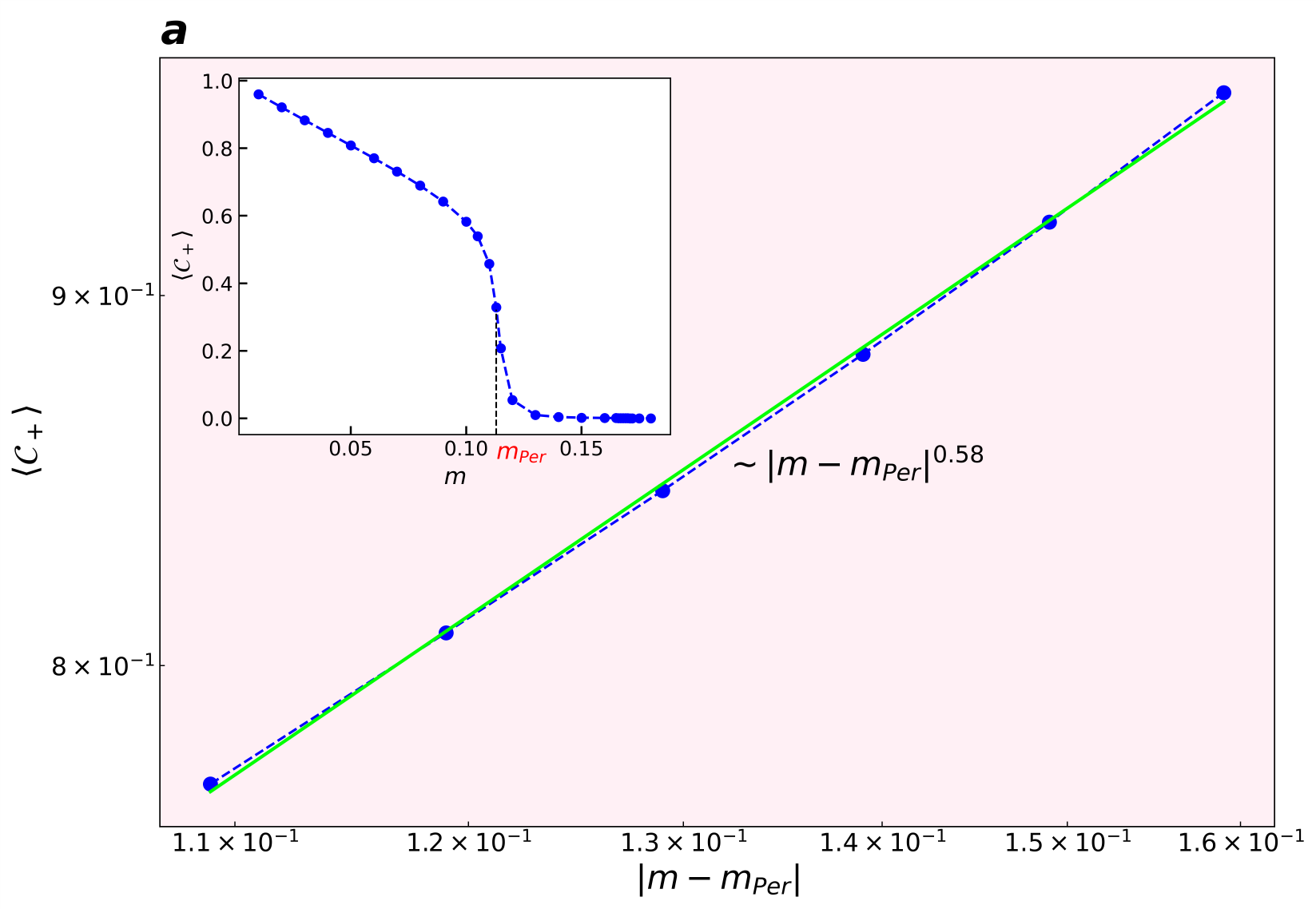}
    \hfill
    \includegraphics[width=\columnwidth]{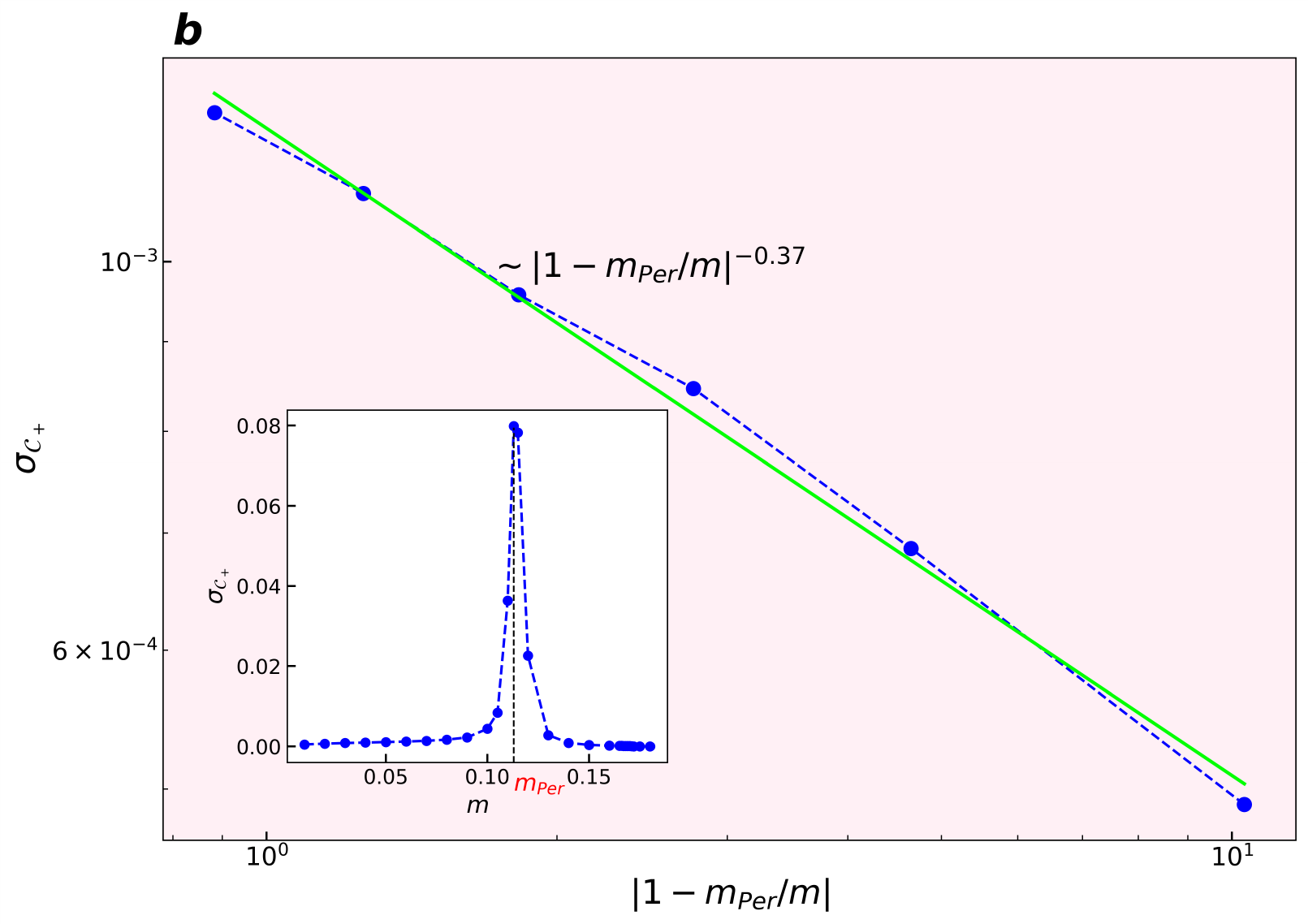}
    \caption{Panel \textbf{(a)} 
        Main: A plot of $\langle \mathcal{C}_+ \rangle$ vs $|m - m_{Per}|$ in log-log scale for $m<m_{Per}$, where the results are plotted for $L=400$ and are largely independent of size effects. The green solid line represents the best-fit line, corresponding to a power law with a slope of $\beta_2=0.58 \pm 0.01$. Inset: The mean of the relative size of the largest cluster $\langle \mathcal{C}_+ \rangle$ as a function of $m$ on a linear scale.
        Panel \textbf{(b)} 
        Main: A plot of $\sigma_{\mathcal{C}_+}$ vs $|1 - m_{Per}/m|$ in log-log scale for $m<m_{Per}$, where the results are plotted for $L=400$ and are largely independent of size effects. The green solid line represents the best-fit line, corresponding to a scaling law with a slope of $-\gamma^{(2)}_{\sigma}=-0.37 \pm 0.01$.Inset: The root-mean-square deviation of the relative size of the largest cluster $\sigma_{\mathcal{C}_+}$ as a function of mortality rate $m$ on a linear scale.}
    \label{figS2}
\end{figure}
Here we outline some additional features of the degradation transitions which take place in the SCA model defined by Eq.s (1)-(4) of the main text of this work.

As outlined in the main text, the degradation transition happens at the critical mortality rate $m_c=0.169$. Here we conduct an analysis of the mean vegetation density $\langle \rho_+ \rangle$ and its standard deviation $\sigma_{\rho_+}$ as functions of the mortality rate $m$. We plot these values in the inset of Fig.~\ref{figS1}\textbf{(a)} ($\langle \rho_+ \rangle$) and \textbf{(b)} ($\sigma_{\rho_+}$). The continuous degradation transition is signaled by $\langle \rho_+ \rangle$ approaching zero and the peak in $\sigma_{\rho_+}$ at $m=m_c$.
Additionally, we show the same quantities as a function of the distance from the critical point $|m - m_c|$ in the respective main plots. We extract the critical exponents by means of power law fits $\langle \rho_+ \rangle\sim  |m - m_c|^{\beta_1}$ and $ \sigma_{\rho_+} \sim |1 - m_c/m|^{\gamma^{(1)}_{\sigma}}$. We obtain for the critical exponents: $\beta_1=0.42 \pm 0.01$ and $\gamma^{(1)}_{\sigma}=-0.29 \pm 0.01$.

We additionally look at the mean of the relative size of the largest cluster $\langle \mathcal{C}_+ \rangle$ and its standard deviation $\sigma_{\mathcal{C}_+}$ as functions of the distance from the percolation threshold $|m - m_{{Per}}|$, with $m_\text{Per}=0.113$. These results are shown in Fig.~\ref{figS2}\textbf{(a)} ($\langle \mathcal{C}_+ \rangle$) and \textbf{(b)} ($\sigma_{\mathcal{C}_+}$). In order to extract the critical exponent for the percolation transition, we consider power law fits, yielding $\langle \mathcal{C}_+ \rangle \sim |m - m_{{Per}}|^{\beta_2}$ and $\sigma_{\mathcal{C}_+} \sim |1 - m_{{Per}}/m|^{\gamma^{(2)}_{\sigma}}$, with critical exponents $\beta_2=0.58 \pm 0.01$ and $\gamma^{(2)}_{\sigma}=-0.37 \pm 0.01$.

This is in agreement with the results of Reference~\cite{kefi2014early}. We should note that the critical exponents deviate from the universal values in the context of the dynamical percolation transition.

\section{Removing lakes from satellite images using adaptive thresholding}\label{Text-S3}

\begin{figure}
    \centering
        \includegraphics[width=0.45\columnwidth]{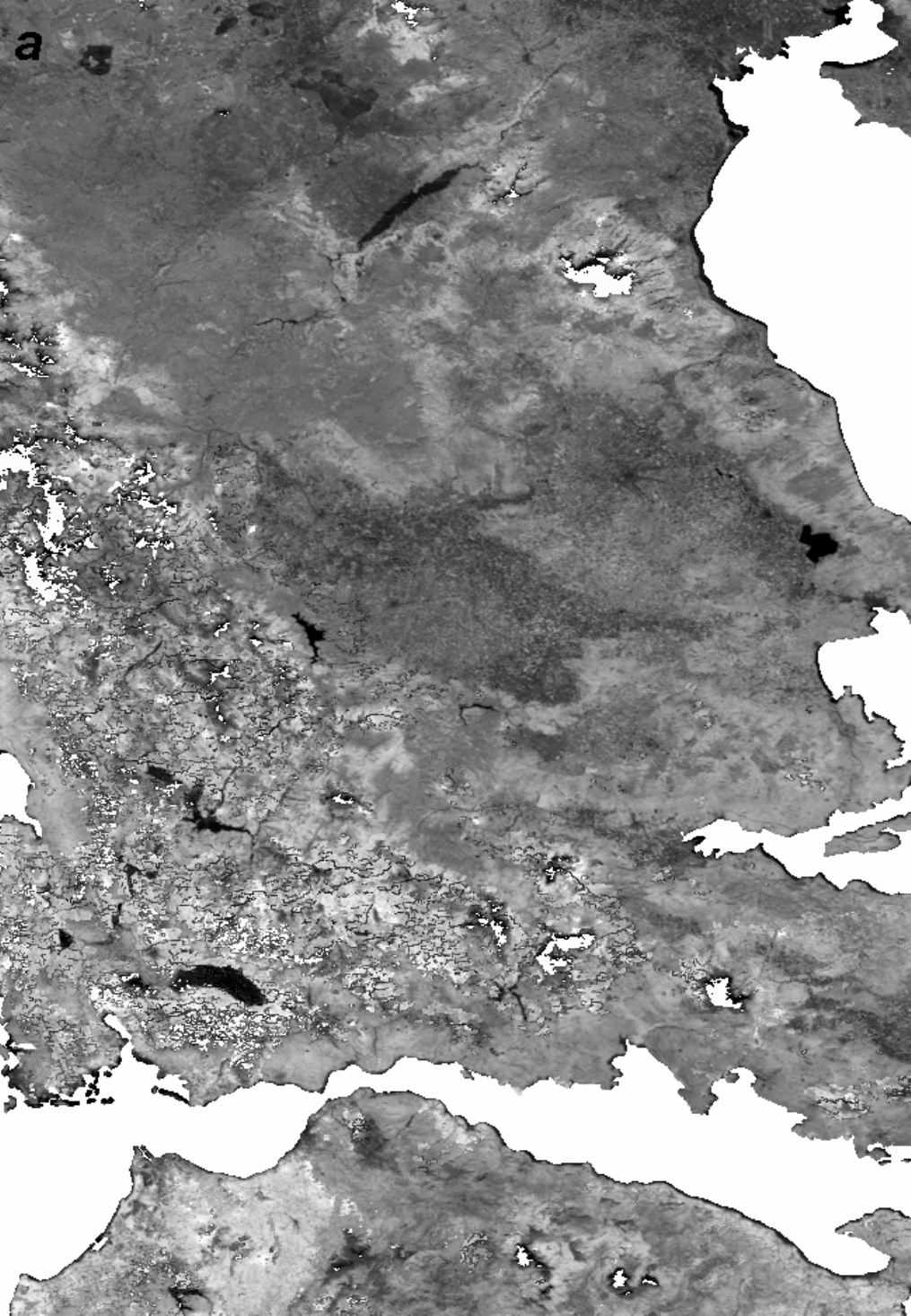}
        $ $
        \includegraphics[width=0.45\columnwidth]{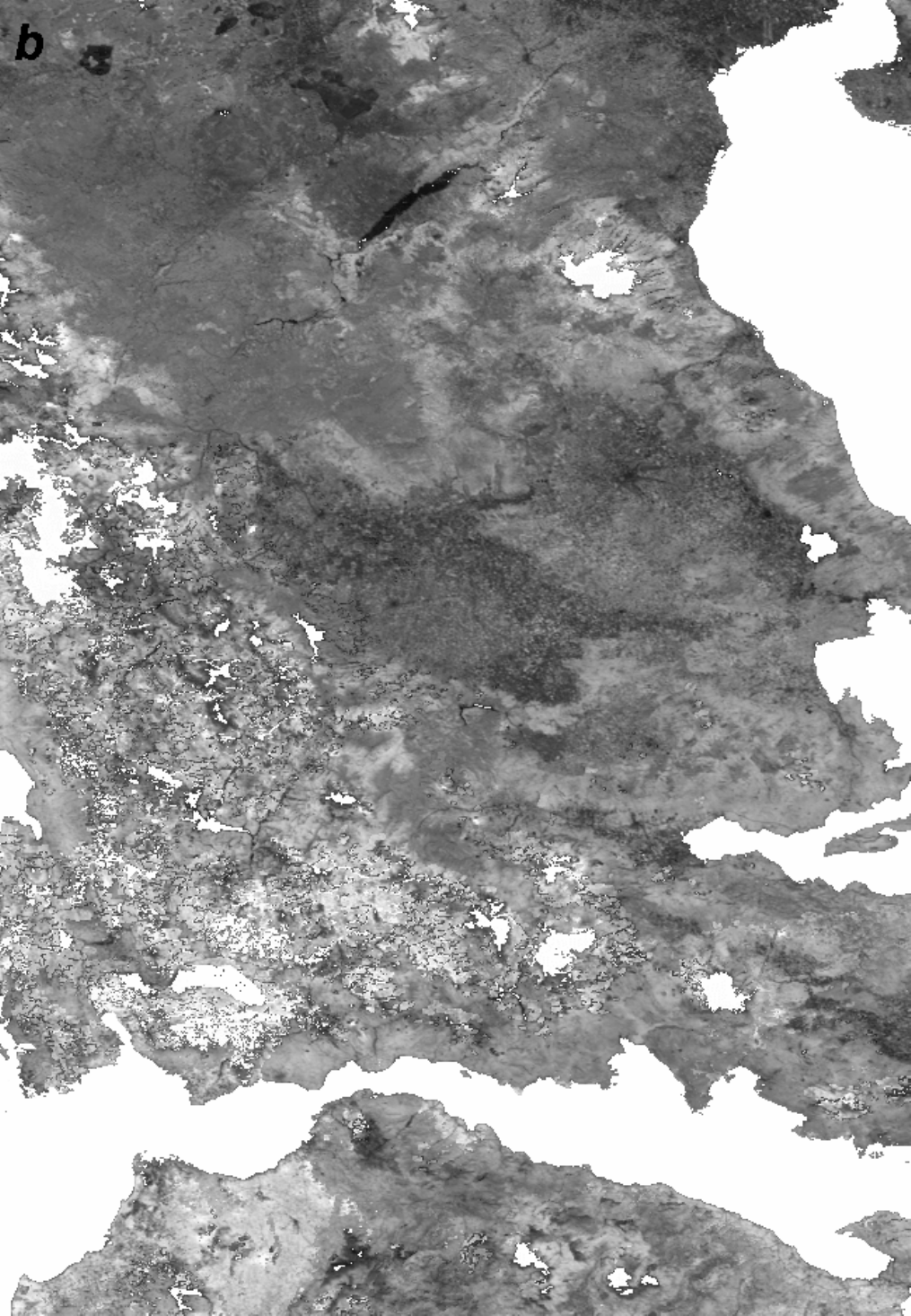}
    \caption{{NDVI-based representation of Greece's vegetation patterns at a specific time.}
    \textit{Left panel: } The original image. \textit{Right panel: } The outcome of applying the threshold-Sauvola method.
    }
    \label{figS3}
\end{figure}

We aim to analyze vegetation dynamics using two indices: the Normalized Difference Vegetation Index (NDVI) \cite{pettorelli2013normalized} and the Leaf Area Index (LAI) \cite{fang2019overview}. To acquire the necessary data, we access the Copernicus Global Land Service~\cite{copernicus} website, which is the European Union's Copernicus Earth observation program database for comprehensive global vegetation products. Within this dataset, the pixel values for the NDVI analysis range from $0$ to $254$. We normalize the pixel values by dividing them by the maximum value observed in the dataset. Importantly, the highest pixel value within our dataset represents regions occupied by seas and oceans (as marked by the database pre-processing), and pixels falling below a certain threshold denote lakes and other bodies of water that have not been automatically removed by the database pre-process. Higher positive values indicate the existence of dense vegetation. In order to obtain a correct sample normalization with respect to the actual land size, we need to exclude water bodies from our analysis. 
The most successful method for identifying the threshold to remove the lakes from our dataset is the Threshold-Sauvola method~\cite{sauvola1997adaptive},  a specialized and local thresholding algorithm. This technique takes into account local variations in image intensity and adaptively adjusts the threshold value accordingly \cite{sauvola1997adaptive}. The processed images depicting NDVI data in Greece on a specific day can be observed in Fig.~\ref{figS3}\textbf{(a)} before applying the method, while Fig.~\ref{figS3}\textbf{(b)} displays the image after the application.

\section{Calculating the numerical curve in the $(\rho, \mathcal{C})$ plane and analyzing the effects of changing parameters}\label{Text-S4}

\begin{figure}
	\centering
	\includegraphics[width=3.5in]{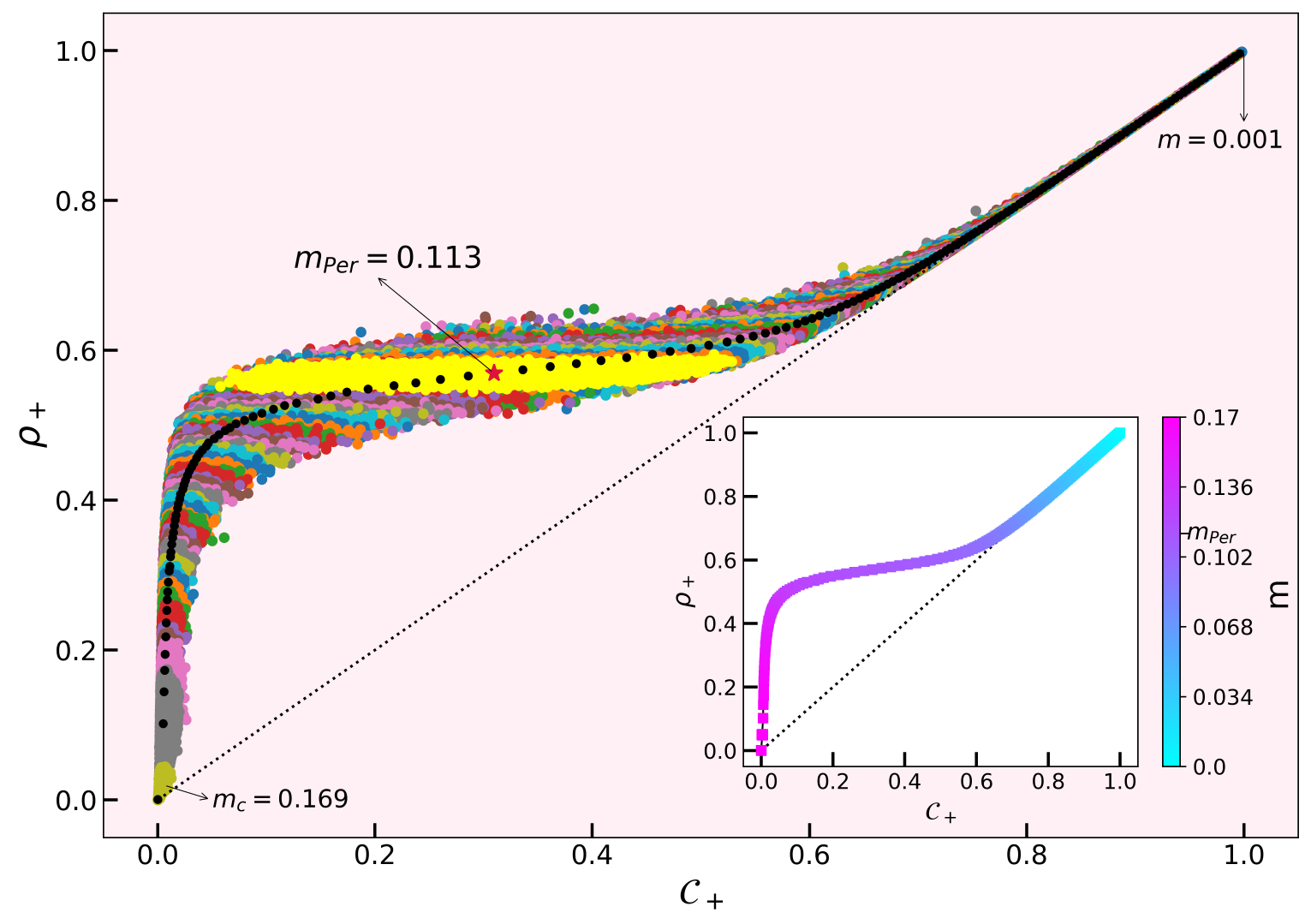}
	\caption{$\rho_+$ vs $\mathcal{C}_+$ for simulated samples of the cellular automata model, for increasing values of the mortality rate $m$ on a lattice of size $L=100$.
        Inset: Centroids of each cloud of points for a given value of $m$. The mortality rate $m$ is color-coded on the $\rho_+$ vs $\mathcal{C}_+$ curve.
        }
	\label{figS4}
\end{figure}

\begin{figure}
	\centering
	\includegraphics[width=3.5in]{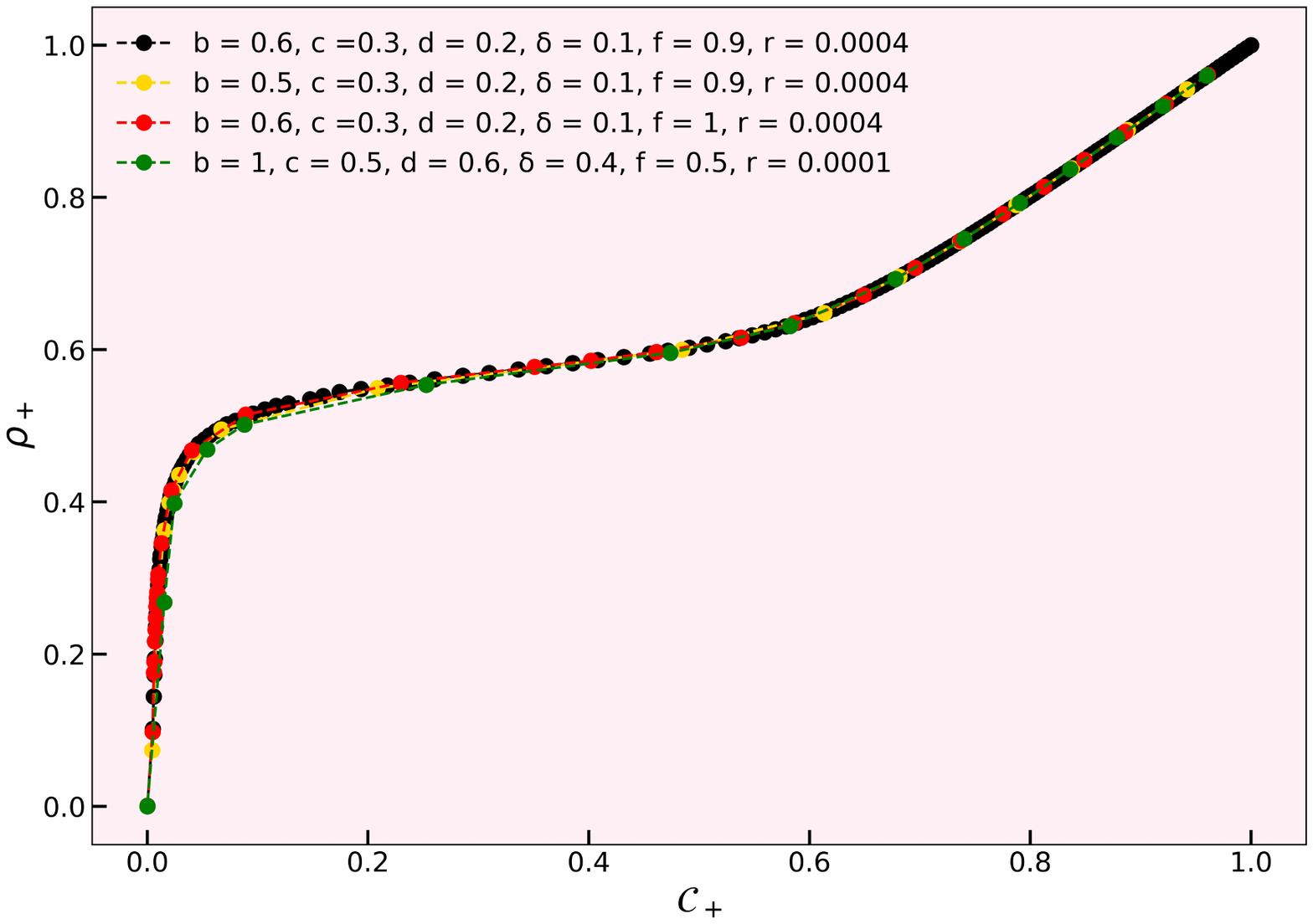}
	\caption{The curve of the $(m, \rho_+, \mathcal{C}_+)$ data points derived from the numerical simulation of the SCA model. The points are color-coded by their distance from $\mathcal \mathcal{C}_+ =0$. The curve displays distinct configurations: The black curve is defined by the parameters $b=0.6, c=0.3, d=0.2, \delta=0.1, f=0.9, r=0.0004$; a yellow curve with $b = 0.5$ and other parameters equal to the black curve; a red curve corresponding to $f = 1$ and other parameters equal to the black curve; and a green curve defined by parameters $b=1, c=0.5, d=0.6, \delta=0.4, f=0.5, r=0.0001$. 
        }
	\label{figS5}
\end{figure}

\begin{table*}[t] 
	\centering 
	\setlength\extrarowheight{3.2pt}
	\caption{\textbf{Geographical coordinates and dimensions of the satellite images for each region} }\label{table1}\label{tab:coordinates}
	
	\begin{tabularx}{\textwidth}{X|X|X|X|X}
		\hline
		
		\rowcolor{ghostwhite}
		\centering  Region
		&\centering Latitude Range  
		&\centering Longitude Range
		&\centering Image Dimensions   
	 \tabularnewline
		
		\colourpadding{black}
		\rowcolor{grannysmithapple}
		\centering France
		&\centering $43.6009$ to $49.2776$
		&\centering $-0.6392$ to $5.8648$ 
		&\centering $1907$ $\times$ $2185$
		\tabularnewline

		\rowcolor{grannysmithapple}
		\centering Germany
		&\centering $50.7715$ to $53.4145$ 
		&\centering $7.3127$ to $13.8169$ 
		&\centering $888$ $\times$ $2185$
		\tabularnewline

		\rowcolor{grannysmithapple}
		\centering Ireland
		&\centering $52.2316$ to $54.0387$
		&\centering $-9.0896$ to -$6.4198$
		&\centering $607$ $\times$ $897$
		\tabularnewline
		
		\colourpadding{black}

		\rowcolor{babypink}
		\centering Spain
		&\centering $37.7445$ to $42.3097$ 
		&\centering $-6.2500$ to $-0.8152$
		&\centering $1534$ $\times$ $1826$
		\tabularnewline

		\rowcolor{babypink}
		\centering Greece
		&\centering $37.8524$ to $40.6447$ 
		&\centering $21.0865$ to $23.0171$ 
		&\centering $938$ $\times$ $649$
		\tabularnewline

		\hline 
		\hline
	\end{tabularx}
\end{table*}

In Fig. 5 of the main text, we show the curve relating the vegetation density $\rho_+$ and the relative size of the largest cluster $\mathcal{C}_+$ with the mortality rates $m$  for the SCA model (1)-(4). This numerical curve is obtained from cloud centroids of the points resulting from several simulations at a given value of $m$. In Fig.~\ref{figS4}, we show $\rho_+$ as a function of $\mathcal{C}_+$ for each simulated sample. This plot depicts a collection of clouds, each corresponding to a value of $m$ denoted by a corresponding color. Within each cloud, there are $10 000$ data points representing distinct memoryless time steps in the numerical simulation. Near the degradation threshold $m_c$ we run the simulation for an increased amount of time steps, $300 000$, to overcome a critical slowing down. As shown in Fig.~\ref{figS4}, for low values of $m$ there is minimal dispersion observed around the clouds. However, as the value of $m$ increases, the fluctuations of $\mathcal{C}_+$ become more pronounced, reaching maximum dispersion at the percolation transition point $m_{Per}=0.113$ (depicted as the large yellow cloud).
For higher values of $m$, the fluctuations gradually decrease until reaching $m_c$ at coordinates $(0,0)$, the point marking the complete absence of vegetation in the system. For better clarity in the plot, a line $x=y$ has been included that serves as a reference, noting that data points cannot exist below this line. The black points represent the centers of the clouds. Additionally, 
in the inset of Fig.~\ref{figS4}, the centroids are presented once again, with a color scale showing the corresponding value of the mortality rate $m$.

To investigate the impact of varying parameters in the SCA model on the $\rho_+$ vs $\mathcal{C}_+$ relationship, we did numerical computations at several different parameter sets; namely, we checked combinations of the following values: $b = 0.3, 0.5, 0.6, 1; f = 0.3, 0.5, 0.9, 1; c = 0.1, 0.3, 0.5; d = 0.2, 0.6, 0.8; \delta = 0.1 0.4, 0.7; r = 0.0001, 0.0004, 0.001$. As an example, in Fig. \ref{figS5}, we show three numerical curves each computed with a different parameter set. The black curve is obtained with the parameters used in the main text of this work, which are $b = 0.6$, $c = 0.3$, $d = 0.2$, $\delta = 0.1$, $f = 0.9$, and $r = 0.0004$. The yellow curve represents the model with the same parameters and different $b$ value ($b=0.5$). The red curve corresponds to the same parameters, and different $f$ value ($f=1$). Finally, the green curve is defined by the parameters $b=1, c=0.5, d=0.6, \delta=0.4, f=0.5, r=0.0001$. We observed that the projection of the curve on the $(\rho_+, \mathcal{C}_+)$ plane remains independent of the variations in the parameters.

\section{Additional Analysis of NDVI and LAI data in all the studied regions}\label{Text-S4}

\subsubsection{Geographical locations of the studied regions}

We include here the geographical coordinates, as well as the dimensions of the satellite images specific to each region under study in Table \ref{tab:coordinates}.

\subsubsection{Analysis of additional regions using NDVI data}

We show here the results of the analysis of areas in Germany, Ireland and Spain, which complement the data for France and Greece presented in the main text.
The thresholding values we used for NDVI are $\lambda_1=0.5$ and $\lambda_2=0.7$ for France, Germany, and Greece, $\lambda_1=0.3$ and $\lambda_2=0.5$ for Spain, and $\lambda_1=0.75$ and $\lambda_2=0.85$ for Ireland. 

\subsubsection{Analysis of the LAI dataset}

Here, we show the results obtained using the LAI data from the Copernicus Land database.

We follow the procedure outlined in Section III of the main text, in an analogous way as for the NDVI data. The data is obtained from the Copernicus Land database for years 2014 to 2020 at a 300m resolution. We preprocess the images to remove water bodies, we discretize the data in 3 states (alive, dead, degraded) and we select samples of 100$\times$100 pixels (corresponding to areas of 30km$\times$30km of size). The thresholding values we used for LAI are $\lambda_1=0.1$ and $\lambda_2=0.175$ for France, Germany, and Greece, $\lambda_1=0.03$ and $\lambda_2=0.1$ for Spain, and $\lambda_1=0.2$ and $\lambda_2=0.25$ for Ireland.

\subsubsection{Results of the NDVI and LAI analysis for all regions}

\begin{figure*}
    \centering
    \includegraphics[width=7.3in]{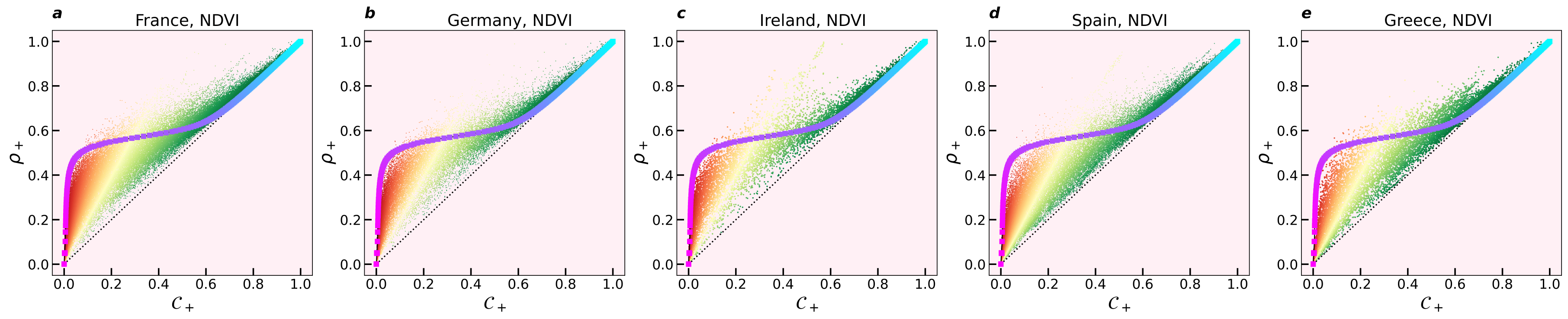}
    \hfill  
    \includegraphics[width=7.3in]{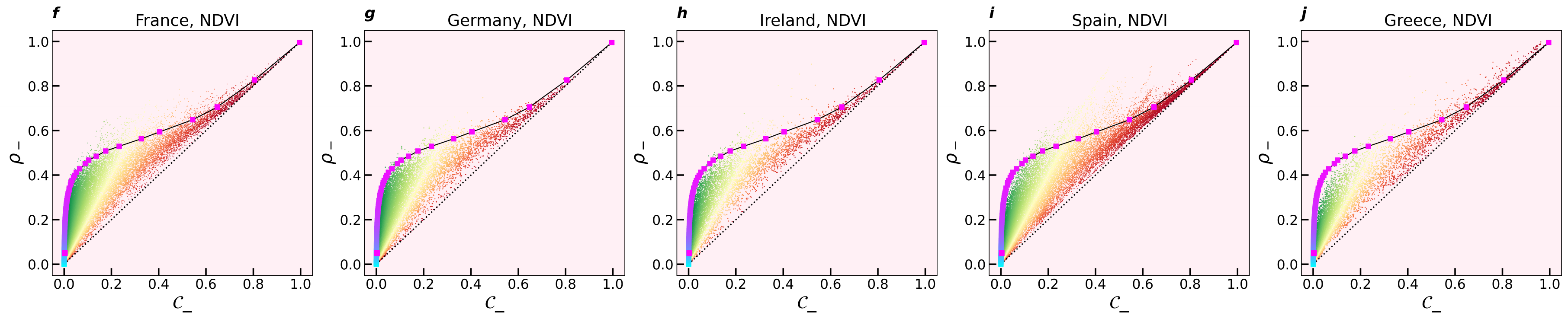}
    \hfill
    \includegraphics[width=7.3in]{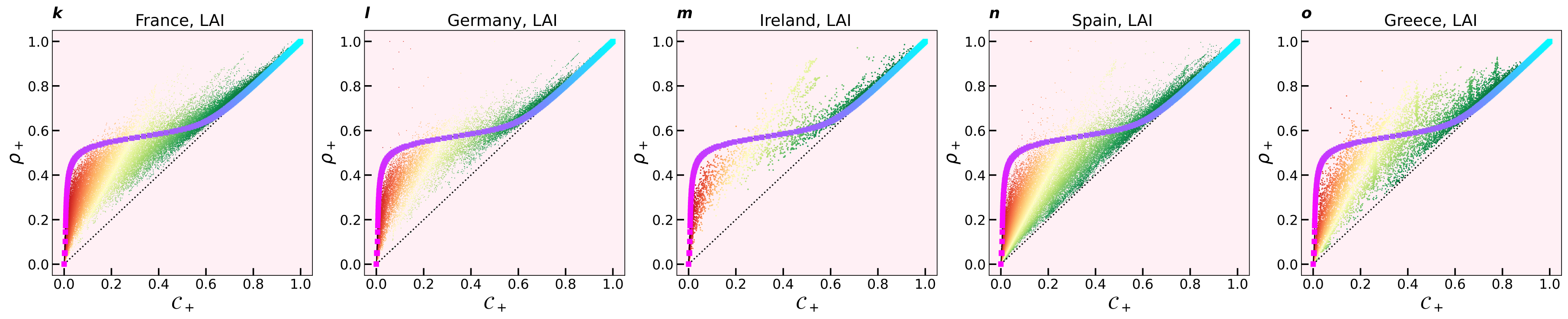}
    \hfill
    \includegraphics[width=7.3in]{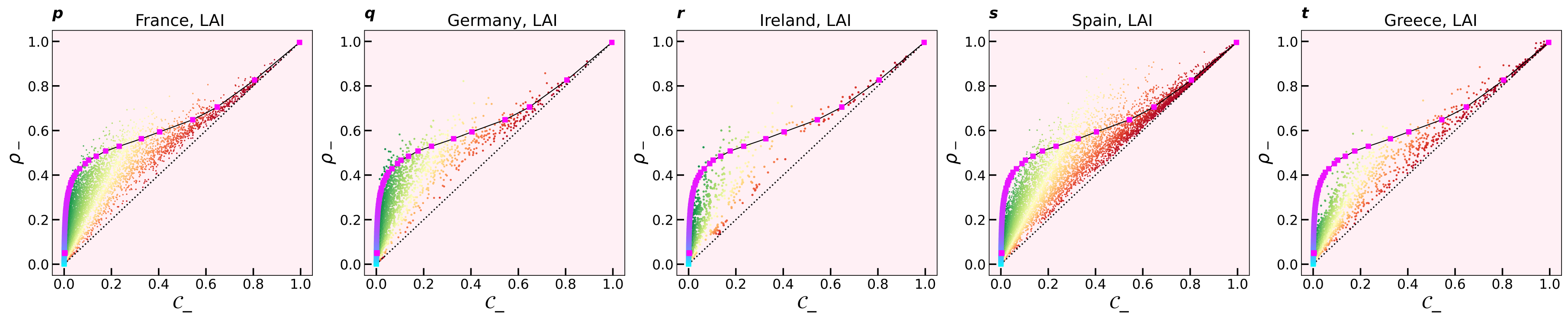}
    \hfill
    \caption{\textbf{(a-e)} {The alive vegetation density $\rho_+$ versus the relative size of the largest cluster of alive cells $\mathcal{C}_+$ and 
    \textbf{(f-j)} {the degraded vegetation density $\rho_-$ versus the relative size of the largest cluster of degraded cells $\mathcal{C}_-$
    for NDVI data across all available dates spanning 2014 to 2020 for all the analyzed regions.}
    \textbf{(k-o)}  $\rho_+$ versus $\mathcal{C}_+$ and \textbf{(p-t)} $\rho_-$ versus $\mathcal{C}_-$ for LAI data across all available dates spanning 2014 to 2020 for all the analyzed regions.}
    }
    \label{figS6}
\end{figure*}

\begin{figure*}
    \centering
    \begin{minipage}[b]{0.29\textwidth}
        \includegraphics[width=\textwidth]{Figs/Fig6a.pdf}
    \end{minipage}
    \begin{minipage}[b]{0.29\textwidth}
        \includegraphics[width=\textwidth]{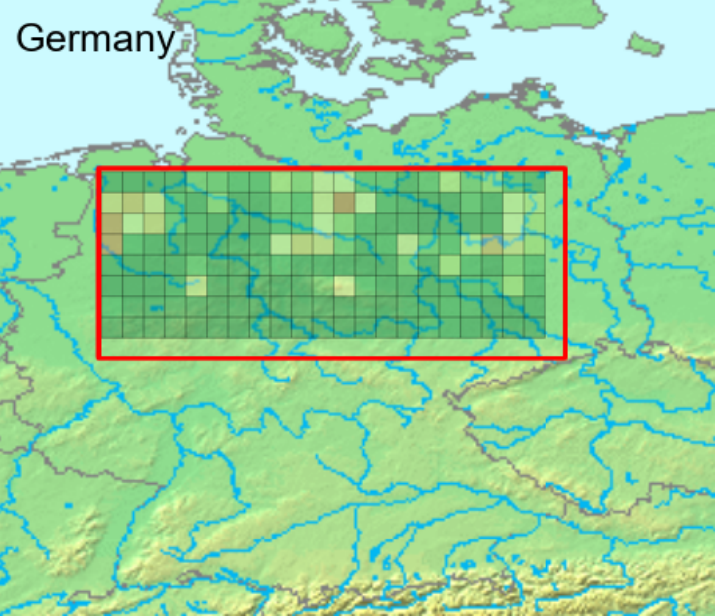}
    \end{minipage}
    \begin{minipage}[b]{0.31\textwidth}
        \includegraphics[width=\textwidth, height=.8\textwidth]{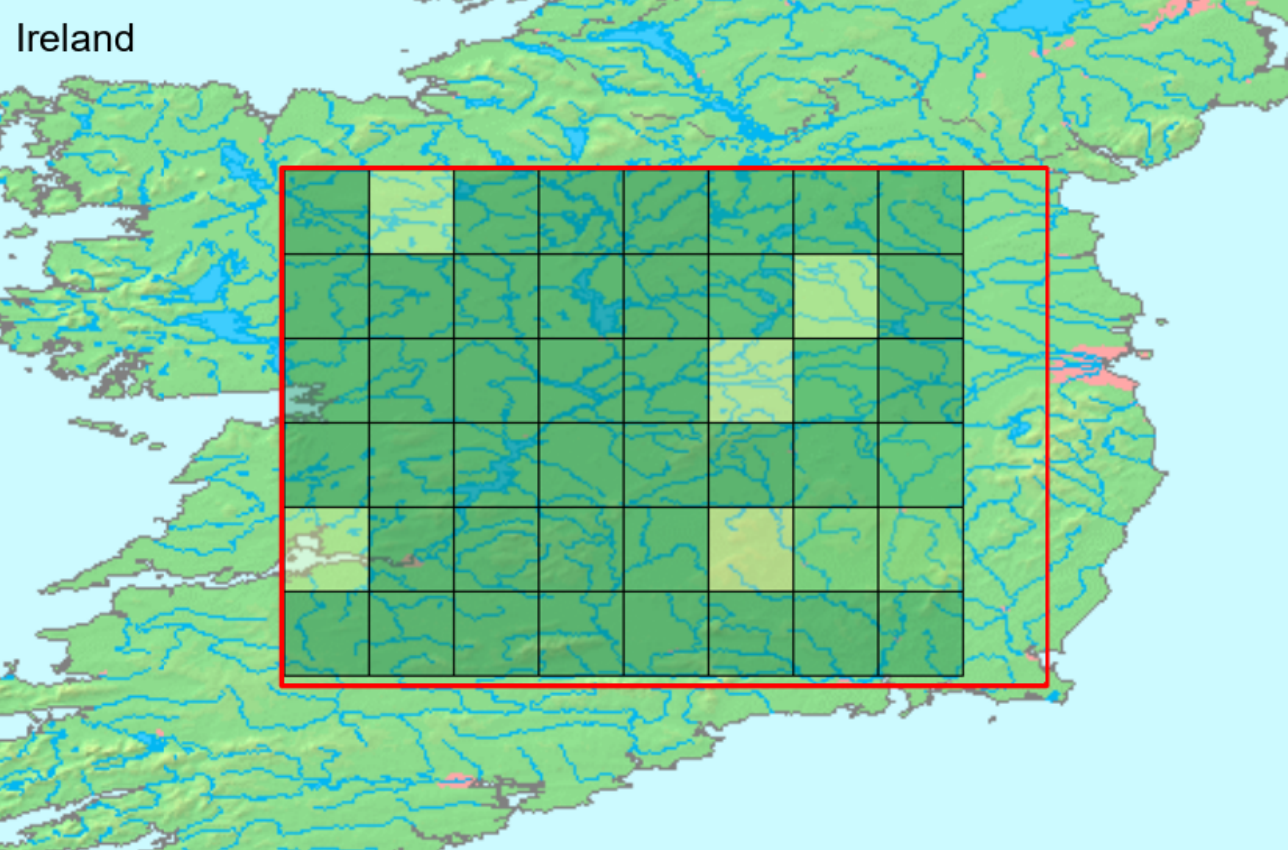}
    \end{minipage}
    \begin{minipage}[b]{0.38\textwidth}
        \includegraphics[width=\textwidth]{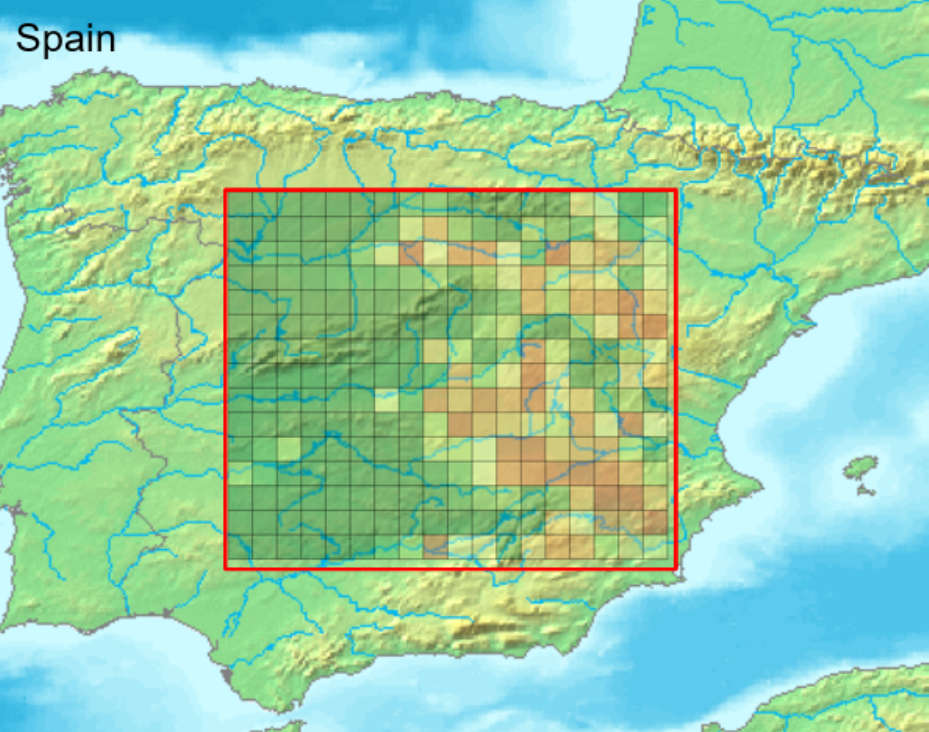}
    \end{minipage}
    \begin{minipage}[b]{0.29\textwidth}
        \includegraphics[width=\textwidth]{Figs/Fig6b.pdf}
    \end{minipage}
    \caption{Qualitative assessment of land degradation, using the value of $ \mathcal{C}_+/\rho_+$ for each sample. The data presented here is obtained from the NDVI dataset on March $11^{th}$ for France, Spain, and Greece, on April $21^{th}$ for Germany, and on June $11^{th}$ for Ireland, all in the year 2020.}
    \label{figS7}
\end{figure*}

\begin{figure*}
    \centering
    \begin{minipage}[b]{0.29\textwidth}
        \includegraphics[width=\textwidth]{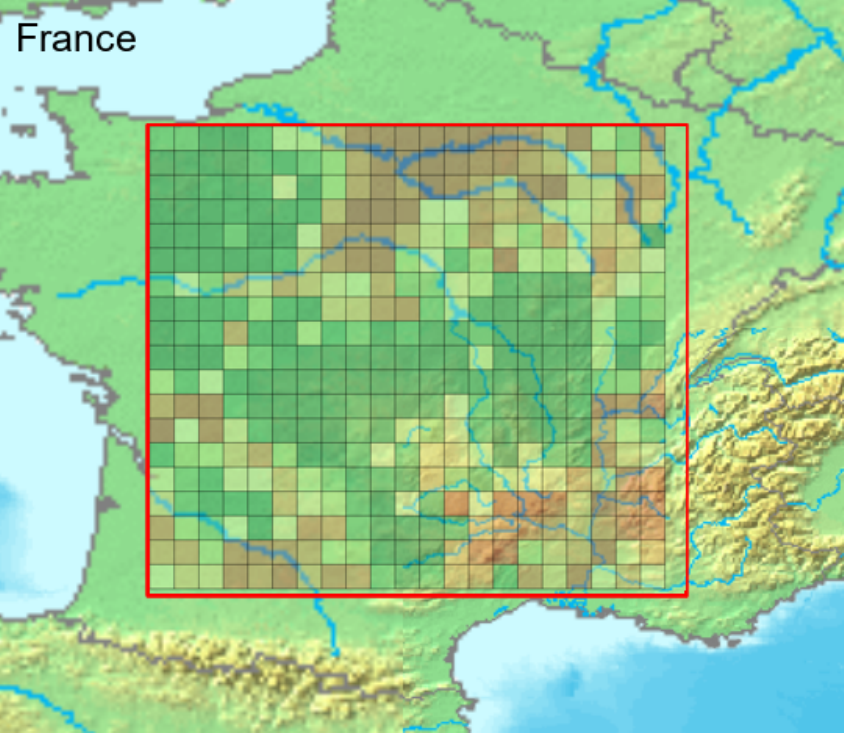}
    \end{minipage}
    \begin{minipage}[b]{0.29\textwidth}
        \includegraphics[width=\textwidth]{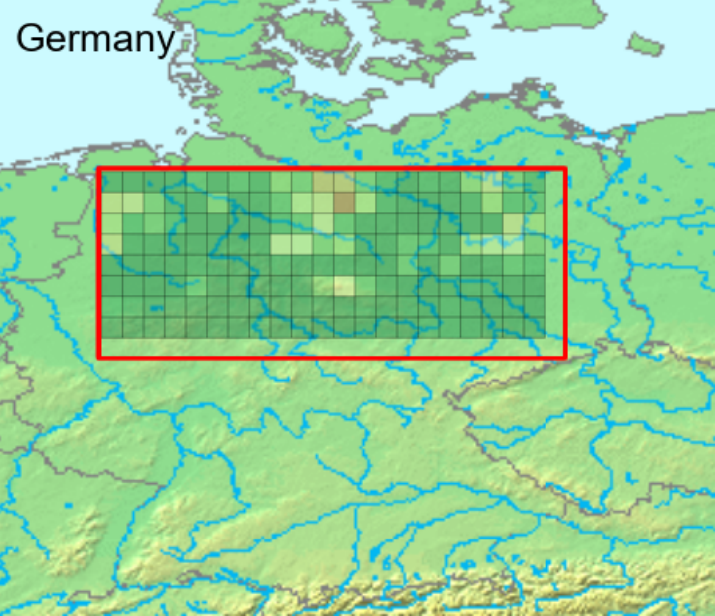}
    \end{minipage}
    \begin{minipage}[b]{0.31\textwidth}
        \includegraphics[width=\textwidth, height=.8\textwidth]{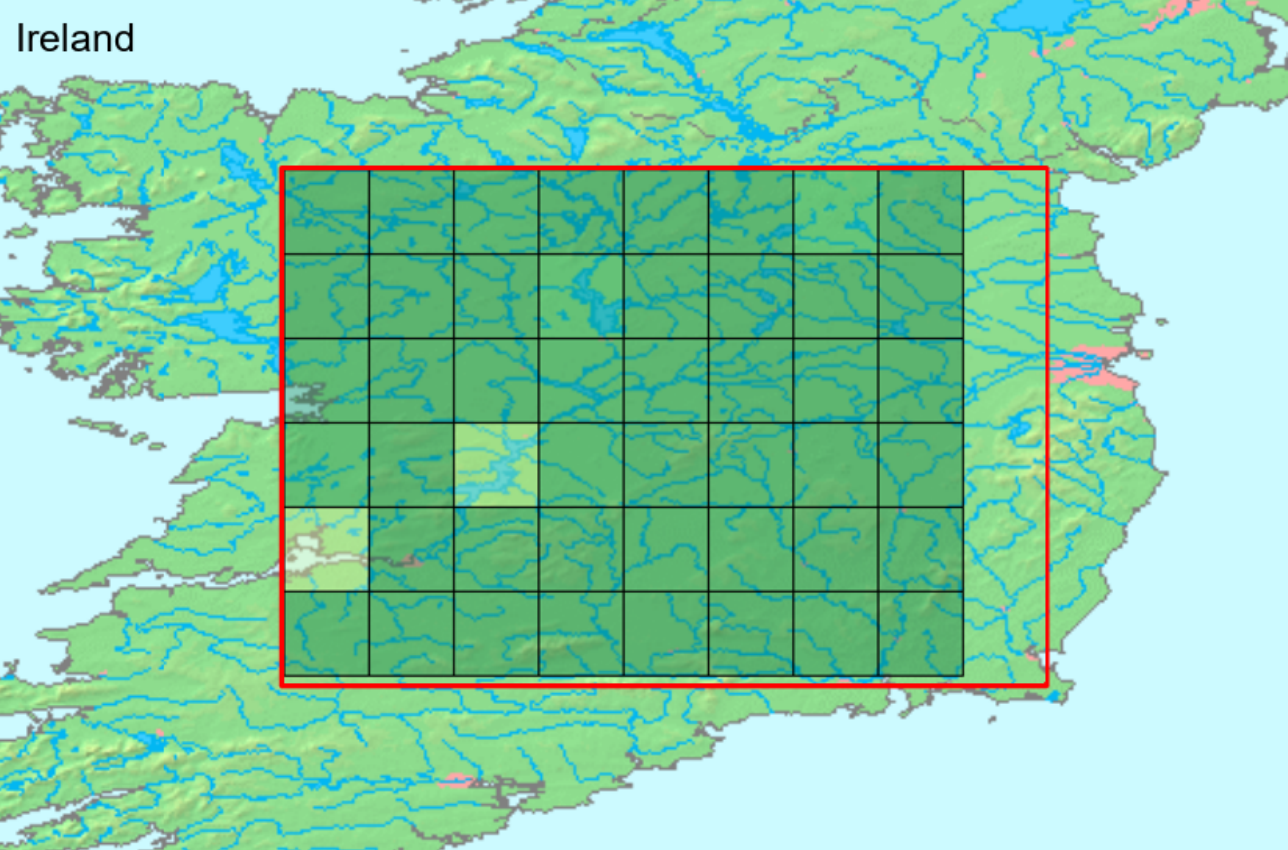}
    \end{minipage}
    \begin{minipage}[b]{0.38\textwidth}
        \includegraphics[width=\textwidth]{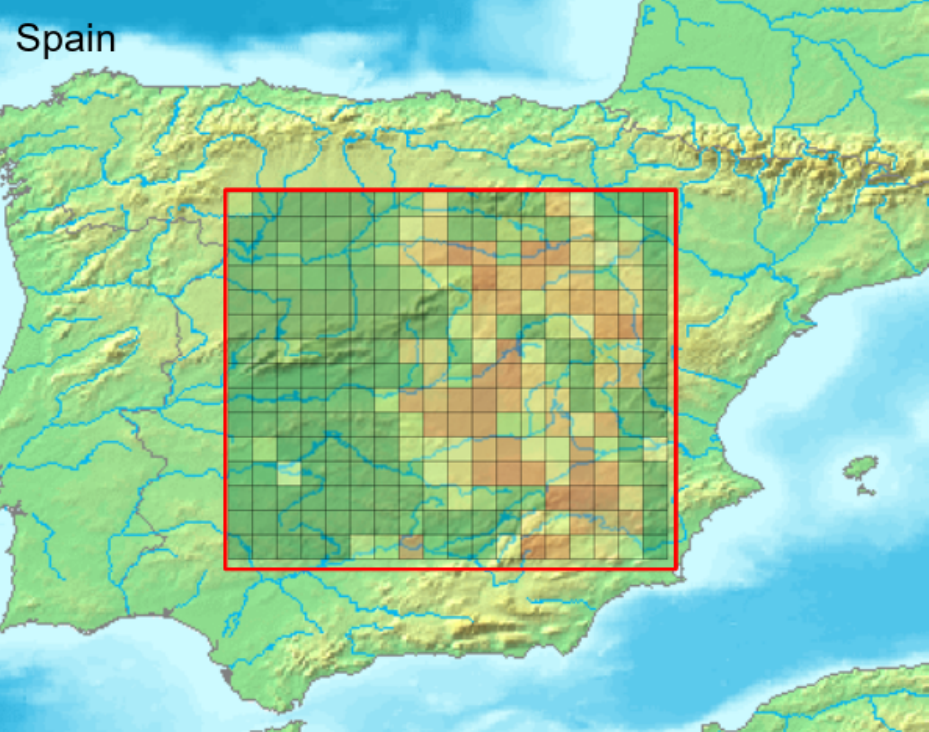}
    \end{minipage}
    \begin{minipage}[b]{0.29\textwidth}
        \includegraphics[width=\textwidth]{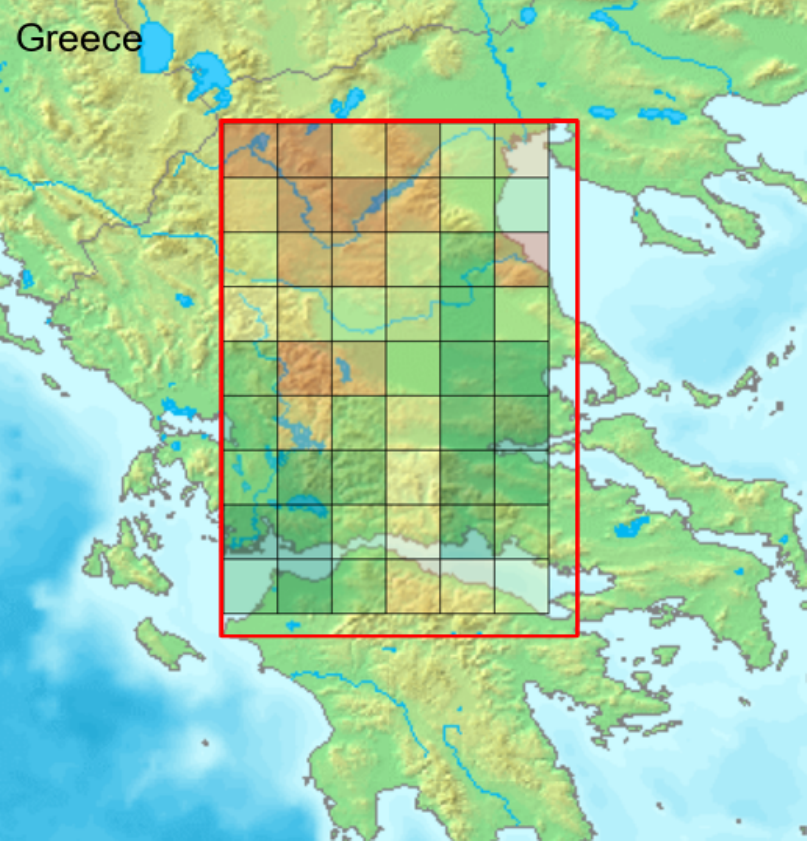}
    \end{minipage}
    \caption{Qualitative assessment of land degradation, using the value of $\mathcal{C}_+/\rho_+$ for each sample. The data presented here is obtained from the LAI dataset on March $10^{th}$ for France, Spain, and Greece, on April $20^{th}$ for Germany, and on June $10^{th}$ for Ireland, all in the year 2020.}
    \label{figS8}
\end{figure*}

Figure~\ref{figS6}\textbf{(a-t)} provide the scatter plots of the alive vegetation density $\rho_+$, degraded vegetation density $\rho_-$, the relative size of the largest cluster of alive cells $\mathcal{C}_+$, and of degraded cells $\mathcal{C}_-$ across various regions, for both NDVI (panels \textbf{(a-j)})  and LAI (panels \textbf{(k-t)})  data. These figures also include a comparison with the SCA  model.
The data is collected on all available dates from $2014$ to $2020$ for NDVI and LAI. Each data point on the graph represents a sub-image measuring $100\times 100$ pixels. The same considerations used in the main text for Fig. 6 (NDVI data for France) are applicable here to qualitatively connect each data point with its corresponding scenario of environmental stress.
Finally, in Figures~\ref{figS7} and~\ref{figS8} we show the sub-images superimposed to a map of the analyzed area, color-coded with the value of $ \mathcal{C}_+/\rho_+$ for each sample for some selected dates in 2020. This qualitatively estimates the vegetation stress for all the analyzed areas in several European countries, for both NDVI and LAI data.

\section{Analysis of the effect of changing the discretization parameters $\lambda_{1,2}$}

The parameters $\lambda_{1,2}$ have been used in the main text in Section IIIA to discretize the satellite image pixel data to the three states of vegetated, empty, and degraded. Their value is set to be below (for $\lambda_1$) and around (for $\lambda_2$) the typical value of the pixel intensity, for both NDVI and LAI data. There is however some degree of arbitrarity in the choice of the precise value, since the type of local vegetation will affect the measured pixel intensity in the satellite data, and particularly as different areas can have different distributions (see the case of Ireland and Spain).

Here, we show how small changes in the definition of the two cutoff values for the discretization affect the subsequent analysis. Namely, we show that the qualitative results identifying degrading areas on a large scale are not affected by the precise choice of the values of $\lambda_{1,2}$, particularly for NDVI data (due to its wider intensity distribution, see Fig. 3 in the main text).

\begin{figure*}
\centering	
\includegraphics[width=0.32\textwidth]{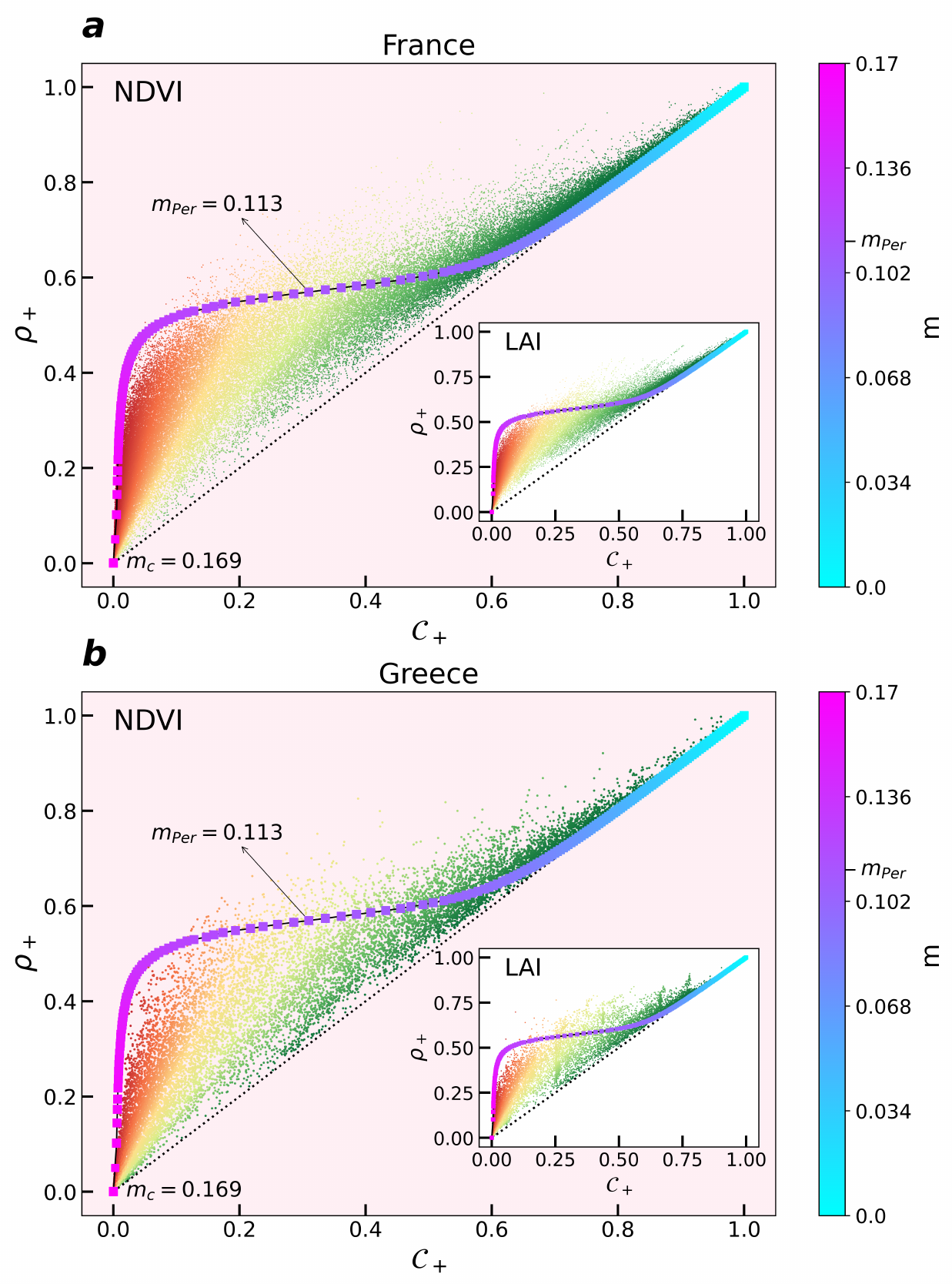}
\includegraphics[width=0.32\textwidth]{Figs/Fig5.pdf}
\includegraphics[width=0.32\textwidth]{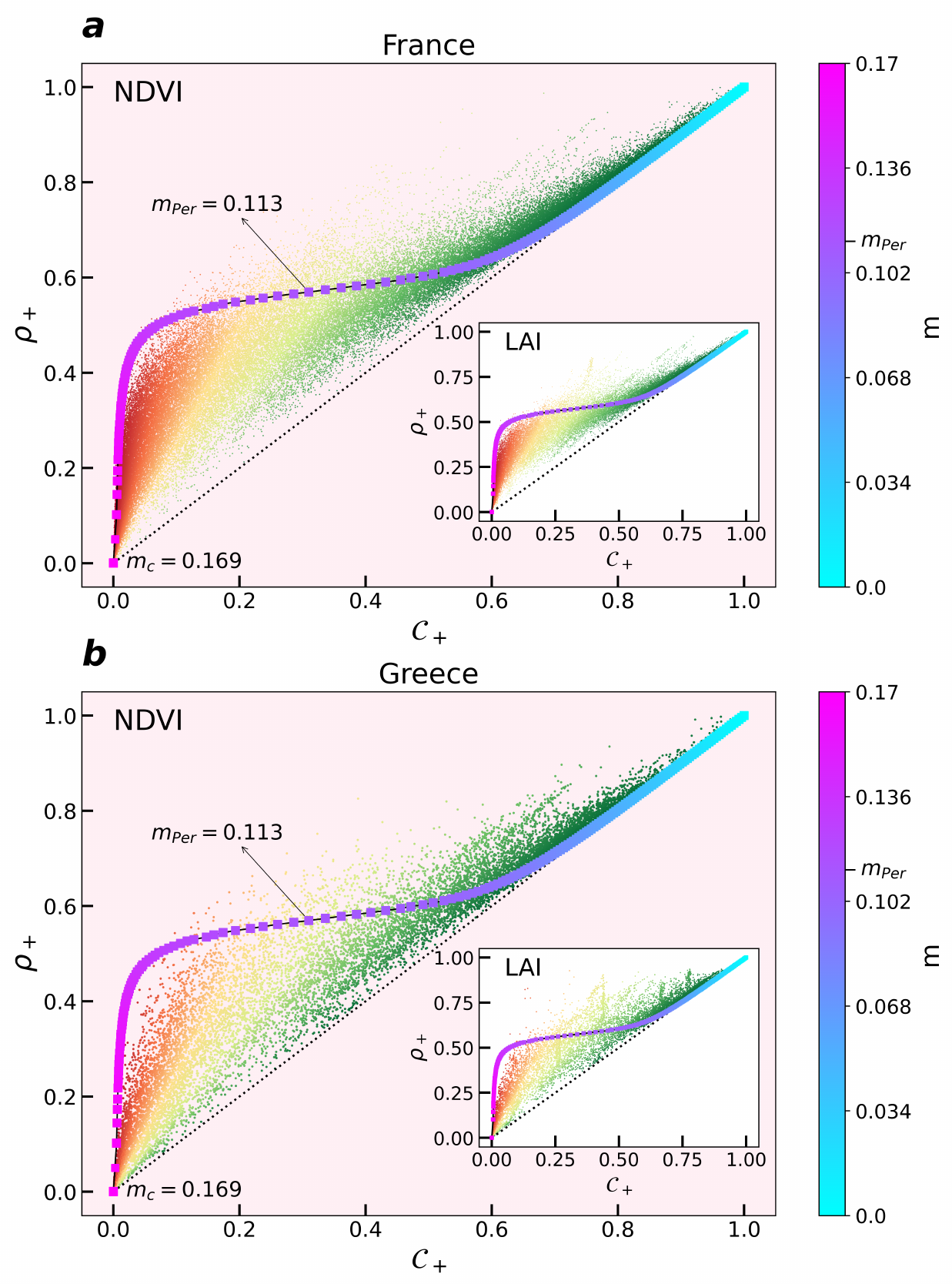}
\caption{Vegetation density $\rho_+$ and the relative size of the largest cluster $\mathcal{C}_+$ for all samples, computed for the cutoff value $\lambda_2=0.7$ used in the main text (center), $\lambda_2 + 0.025$ (left), and $\lambda_2 - 0.025$ (right). We refer to Figure 5 of the main text, which this figure reproduces, for a full description.
 }
	\label{FigS9}
\end{figure*}

In Fig.~\ref{FigS9}, we show a scatter plot of $\rho_+$ and $\mathcal C_+$ analogous to Figure 5 of the main text, recomputed with varying values of the vegetated to empty cutoff $\lambda_2$: we consider values of $\lambda_2\pm 0.025$.
It is evident that the consistency between the model and satellite data does not depend on the definition of the thresholds.

\begin{figure*}
    \centering
    \begin{minipage}[b]{0.28\textwidth}
        \includegraphics[width=\textwidth]{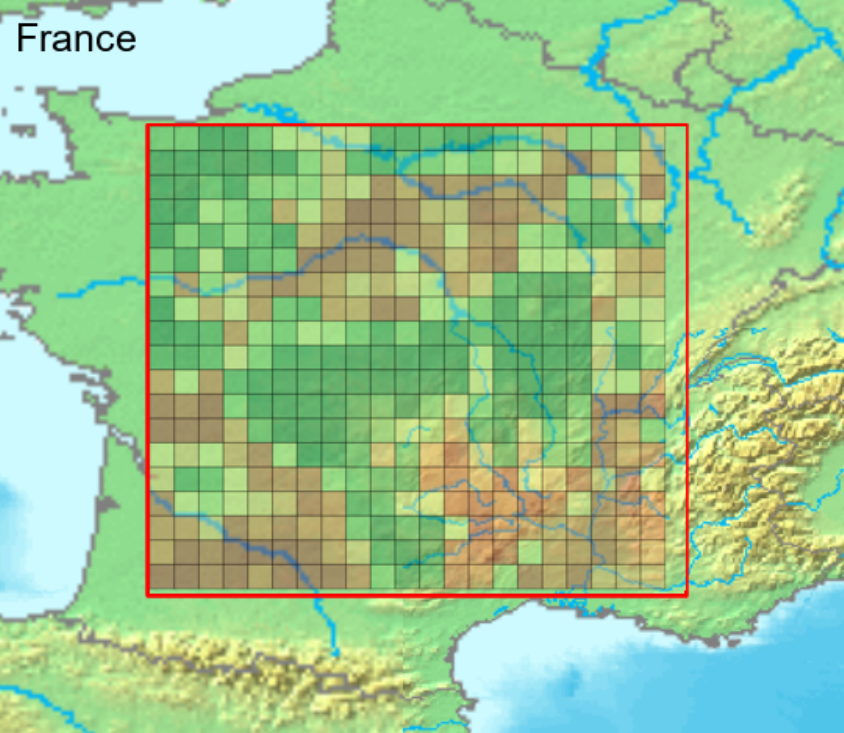}
    \end{minipage}
    \begin{minipage}[b]{0.28\textwidth}
        \includegraphics[width=\textwidth]{Figs/Fig6a.pdf}
    \end{minipage}
    \begin{minipage}[b]{0.28\textwidth}
        \includegraphics[width=\textwidth]{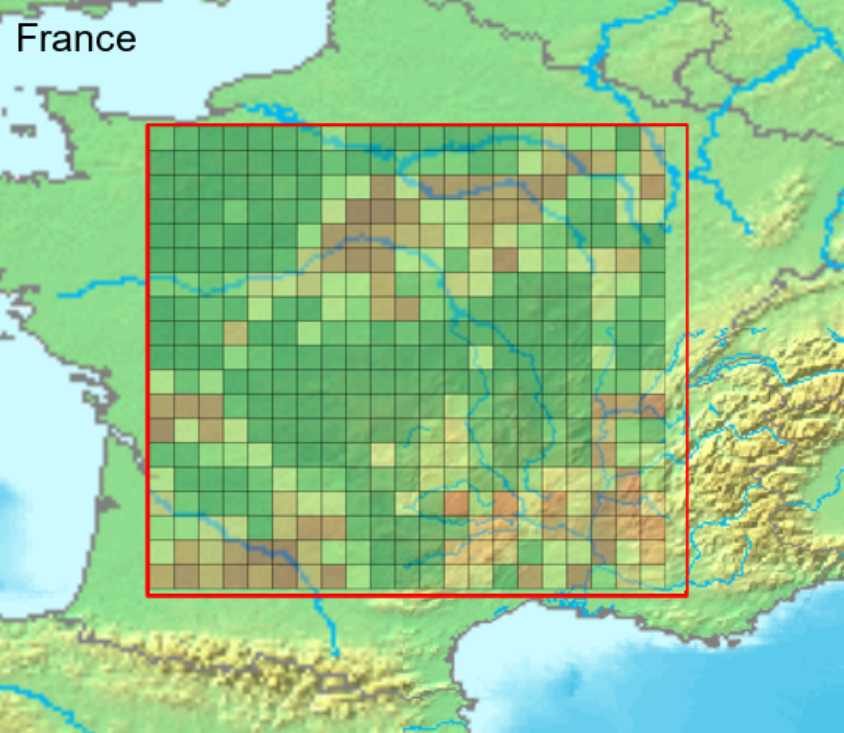}
    \end{minipage}
    \begin{minipage}[b]{0.28\textwidth}
        \includegraphics[width=\textwidth]{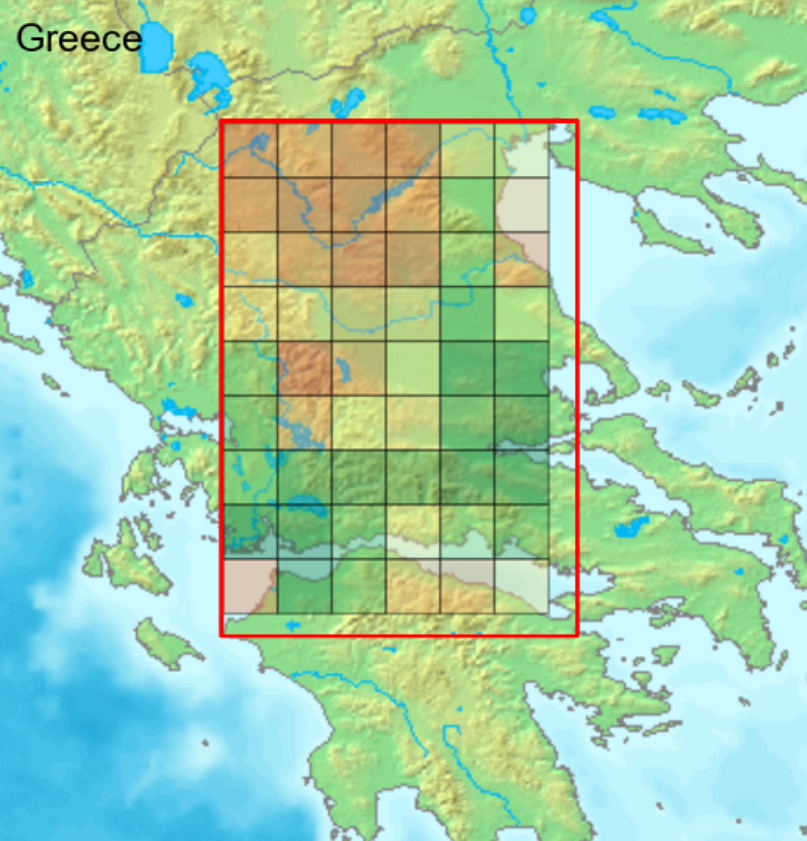}
    \end{minipage}
    \begin{minipage}[b]{0.28\textwidth}
        \includegraphics[width=\textwidth]{Figs/Fig6b.pdf}
    \end{minipage}
    \begin{minipage}[b]{0.28\textwidth}
        \includegraphics[width=\textwidth]{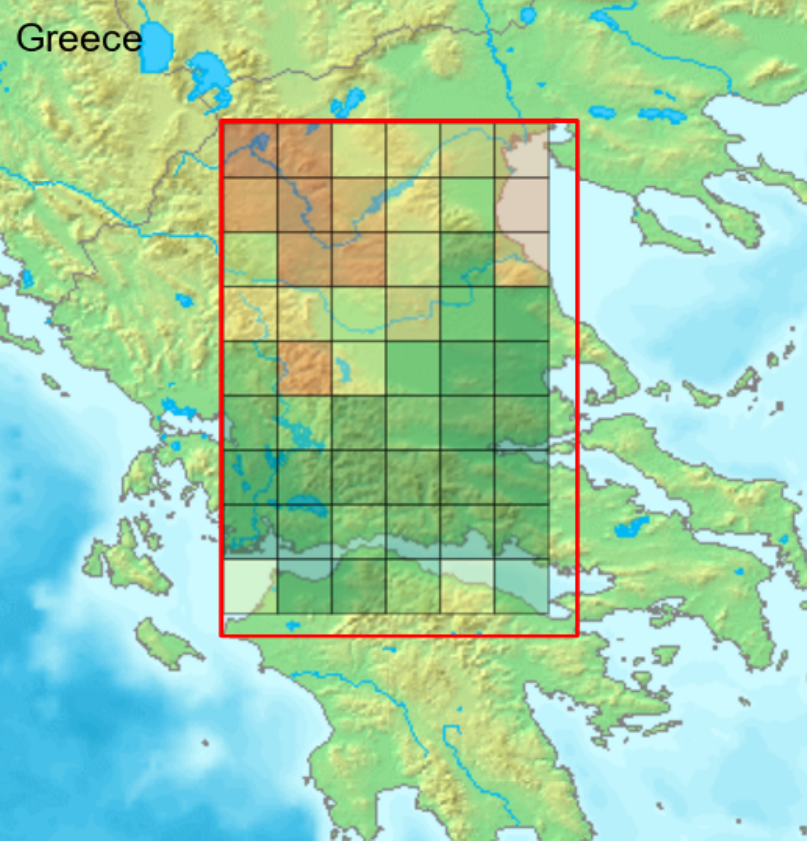}
    \end{minipage}
    \caption{Qualitative assessment of land degradation of France (top row) and Greece (bottom row), from NDVI data obtained for the 11th March 2020. We show a comparison of three values of the vegetated to empty cutoff value: $\lambda_2=0.7$ as used in the main text (center), $\lambda_2 + 0.025$ (left), and $\lambda_2 - 0.025$ (right).}
    \label{figS10}
\end{figure*}

\begin{figure*}
    \centering
    \begin{minipage}[b]{0.28\textwidth}
        \includegraphics[width=\textwidth]{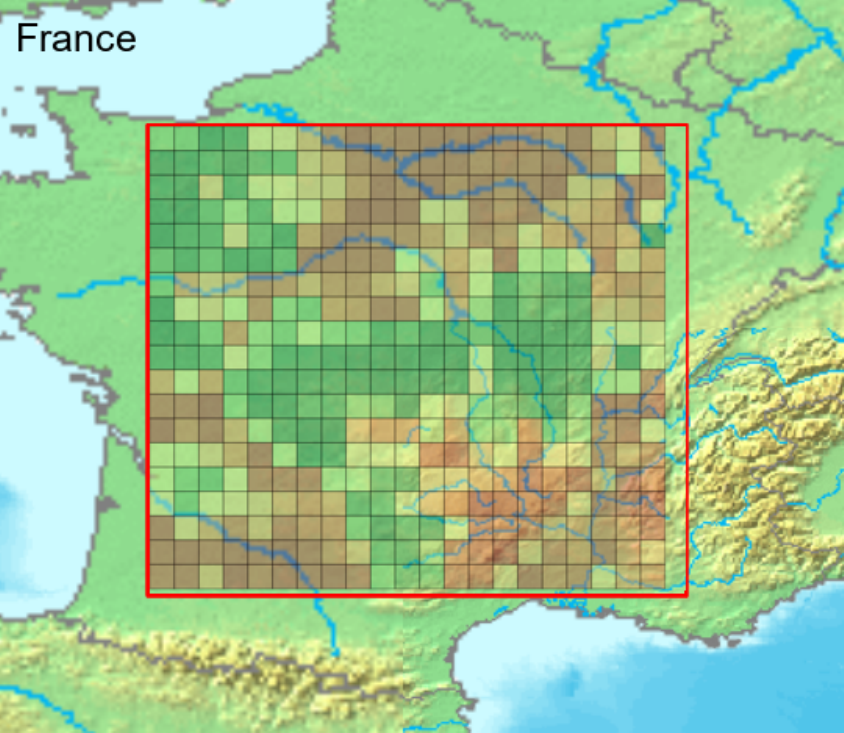}
    \end{minipage}
    \begin{minipage}[b]{0.28\textwidth}
        \includegraphics[width=\textwidth]{Figs/Fig-S8a.pdf}
    \end{minipage}
    \begin{minipage}[b]{0.28\textwidth}
        \includegraphics[width=\textwidth]{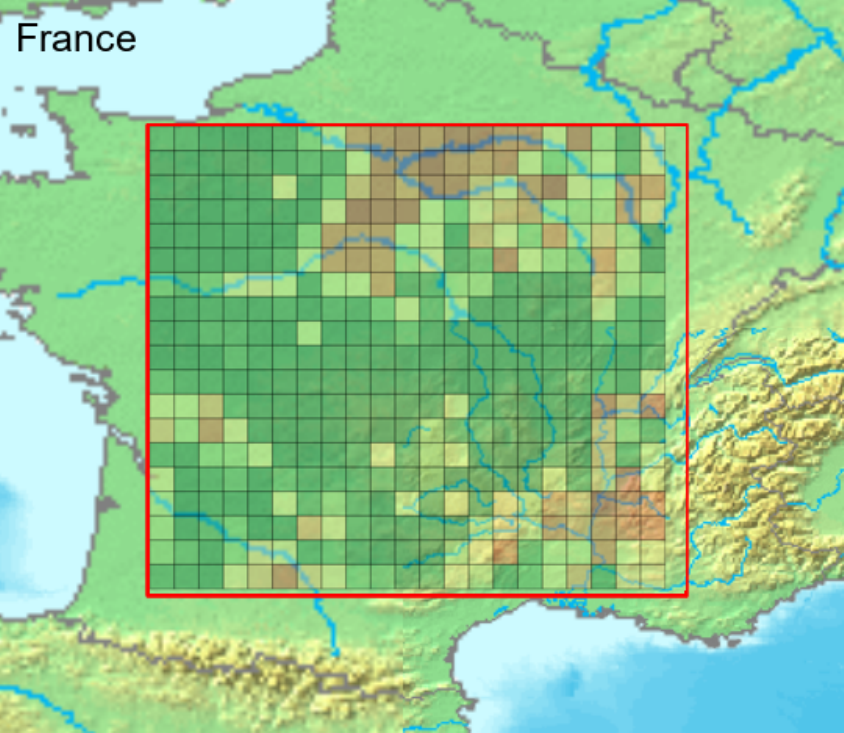}
    \end{minipage}
    \begin{minipage}[b]{0.28\textwidth}
        \includegraphics[width=\textwidth]{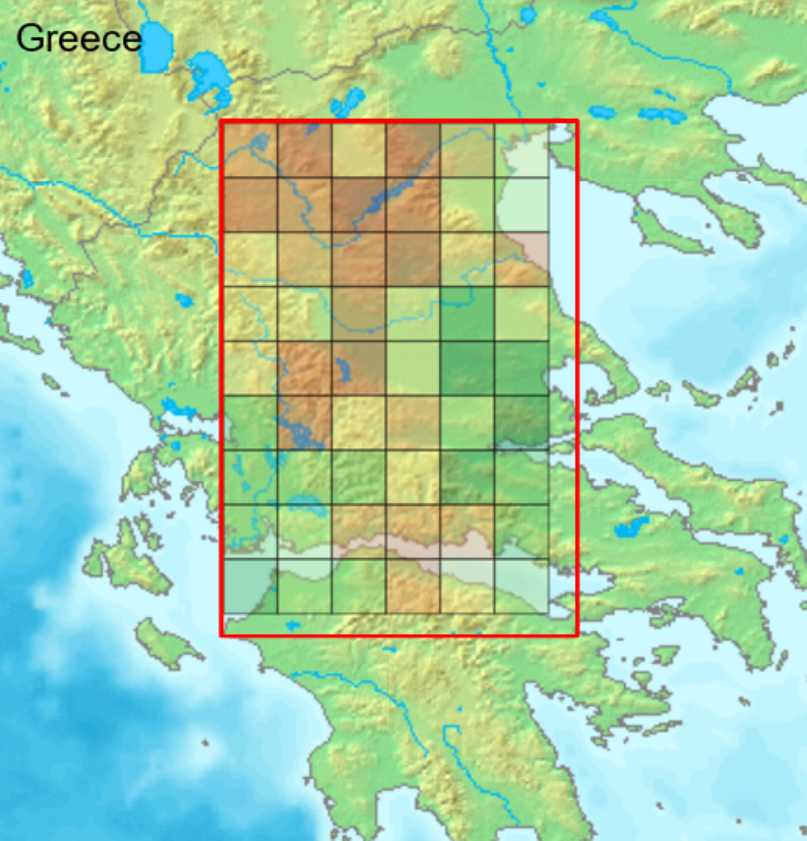}
    \end{minipage}
    \begin{minipage}[b]{0.28\textwidth}
        \includegraphics[width=\textwidth]{Figs/Fig-S8e.pdf}
    \end{minipage}
    \begin{minipage}[b]{0.28\textwidth}
        \includegraphics[width=\textwidth]{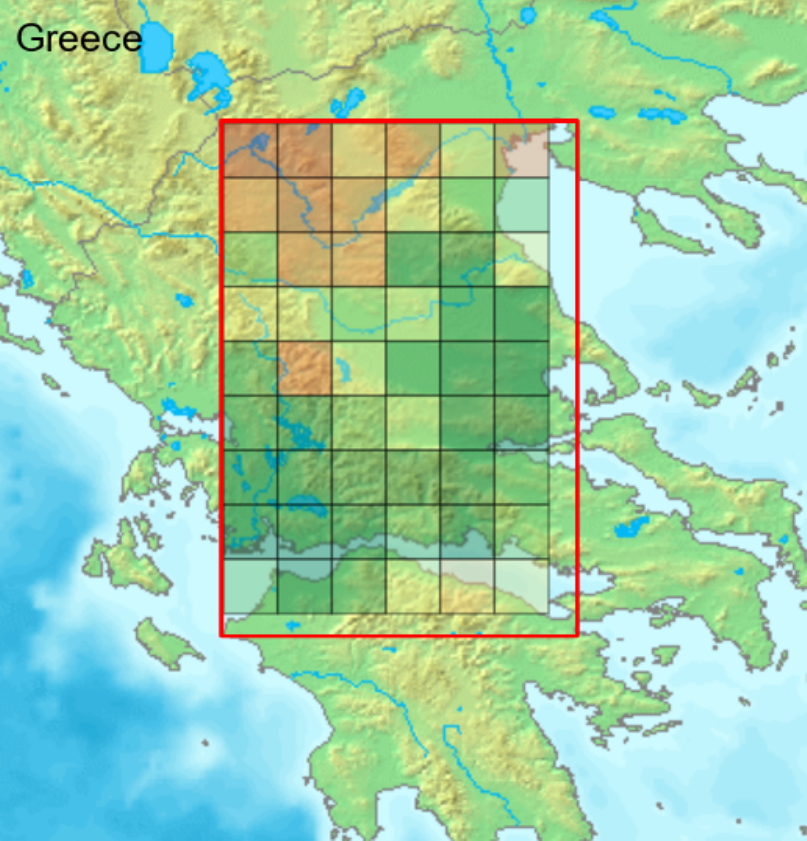}
    \end{minipage}
    \caption{Qualitative assessment of land degradation of France (top row) and Greece (bottom row), from LAI data obtained for the 11th March 2020. We show a comparison of three values of the vegetated to empty cutoff value: $\lambda_2=0.175$ as used in the main text (center), $\lambda_2 + 0.025$ (left), and $\lambda_2 - 0.025$ (right).}
    \label{figS11}
\end{figure*}

Additionally, we have also recomputed the qualitative assessment of land degradation as in Figure 6 of the main text. Specifically, we show the maps obtained from the NDVI and LAI datasets for France and Greece on a specific date, while considering varying values of the threshold, in Fig.s~\ref{figS10} (NDVI) and \ref{figS11} (LAI). 
In particular, we illustrate how these changes in the values do not affect the results.

\section{Tracking of vegetation dynamics}

\begin{figure*}
    \centering
    \begin{subfigure}[b]{0.245\textwidth}
        \includegraphics[width=\textwidth]{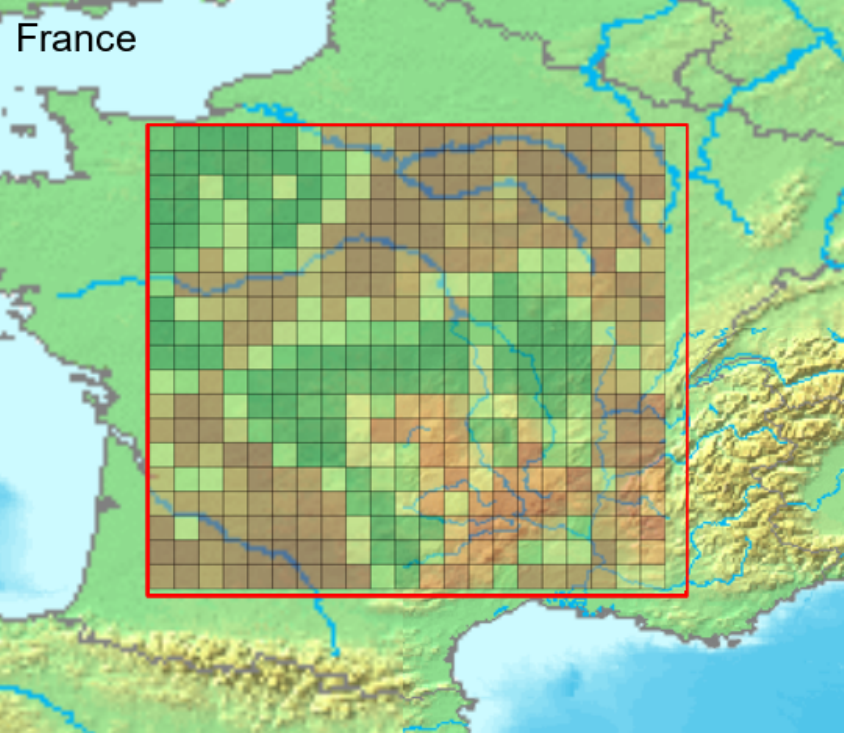}
        \caption{2014}
        \label{fig:a}
    \end{subfigure}
    \begin{subfigure}[b]{0.245\textwidth}
        \includegraphics[width=\textwidth]{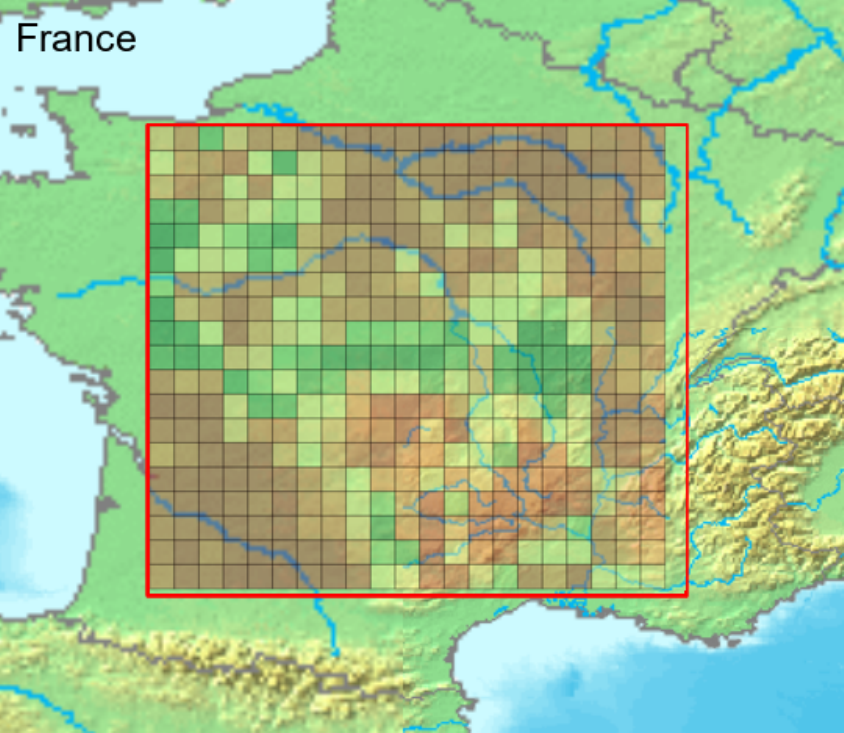}
        \caption{2015}
        \label{fig:b}
    \end{subfigure}
    \begin{subfigure}[b]{0.245\textwidth}
        \includegraphics[width=\textwidth]{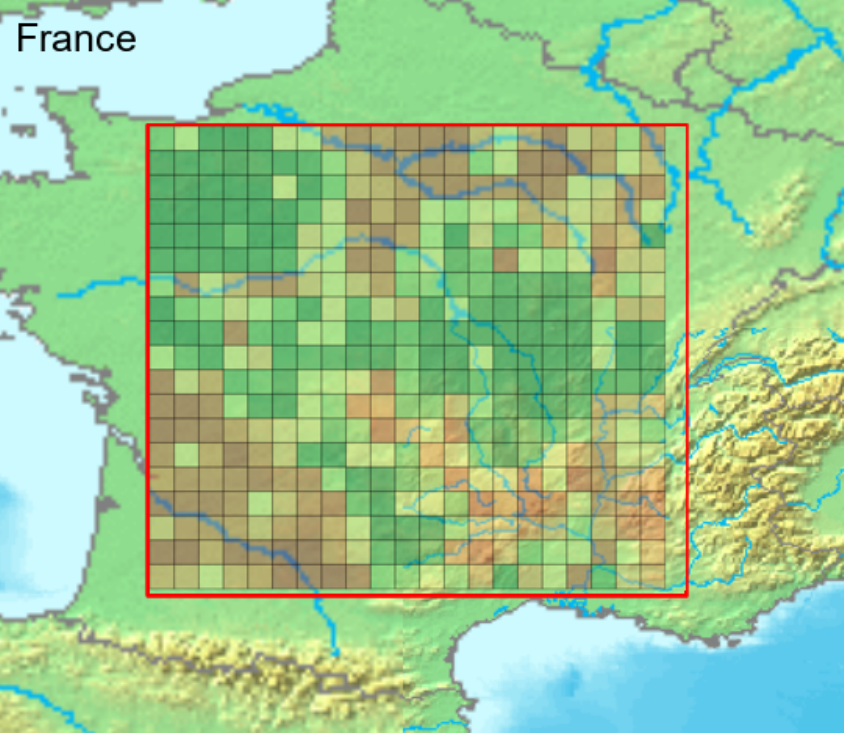}
        \caption{2016}
        \label{fig:c}
    \end{subfigure}
    \begin{subfigure}[b]{0.245\textwidth}
        \includegraphics[width=\textwidth]{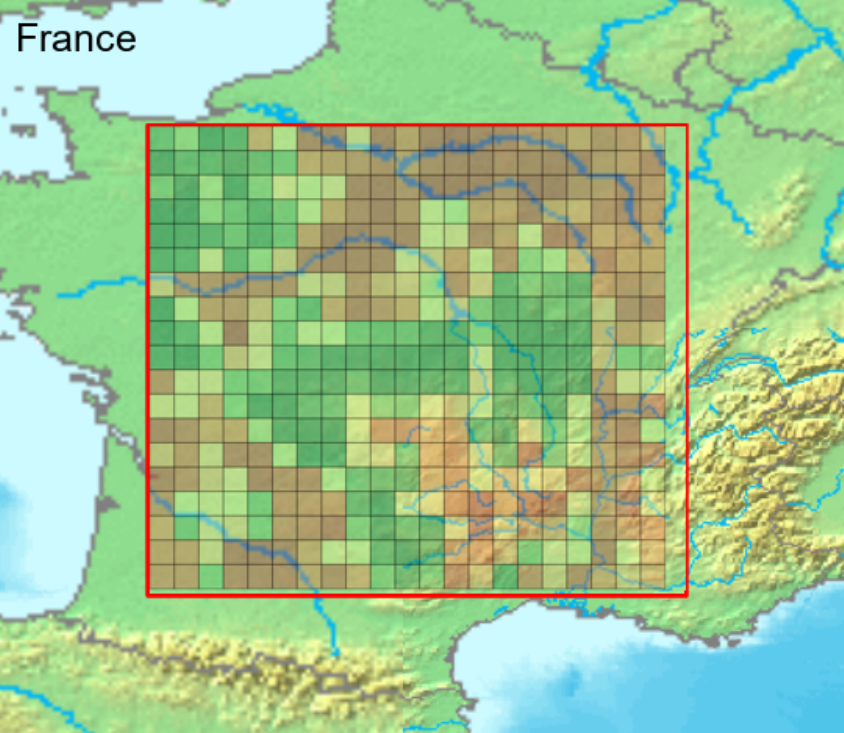}
        \caption{2017}
        \label{fig:d}
    \end{subfigure}
    \begin{subfigure}[b]{0.245\textwidth}
        \includegraphics[width=\textwidth]{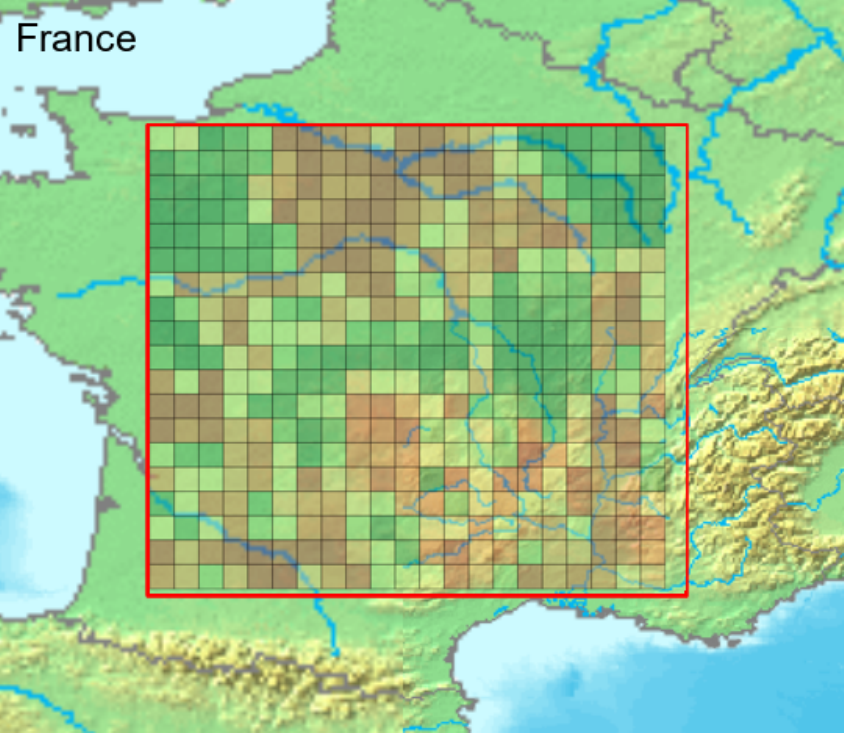}
        \caption{2018}
        \label{fig:e}
    \end{subfigure}
    \begin{subfigure}[b]{0.245\textwidth}
        \includegraphics[width=\textwidth]{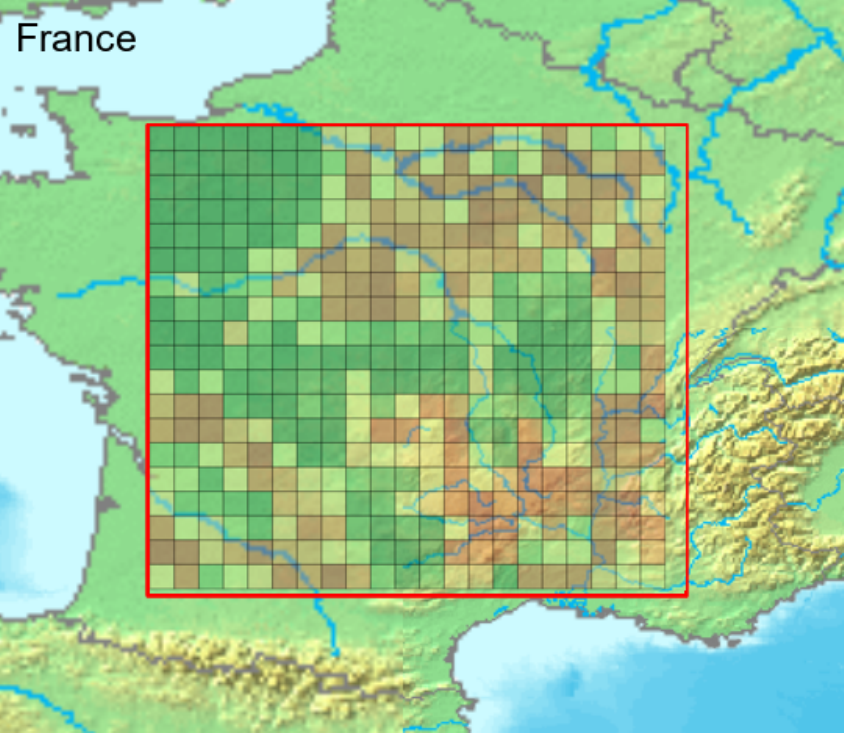}
        \caption{2019}
        \label{fig:f}
    \end{subfigure}
    \begin{subfigure}[b]{0.245\textwidth}
        \includegraphics[width=\textwidth]{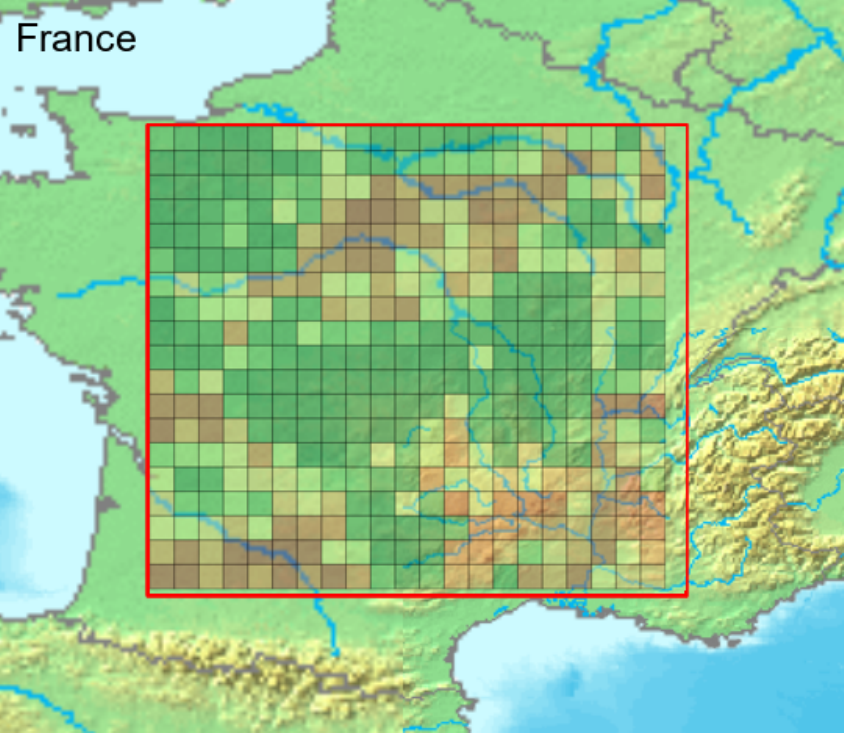}
        \caption{2020}
        \label{fig:g}
    \end{subfigure}
    \caption{Qualitative assessment of land degradation using the value of $ \mathcal{C}_+/\rho_+$ derived from NDVI data on March 11th for France, spanning 2014-2020.}
    \label{figS12}
\end{figure*}

In the subsequent analysis, we illustrate how the method outlined in our work can enhance comprehensive country-scale land monitoring.
Namely, by mapping the qualitative assessment of land degradation year-by-year, long term changes in the vegetation can be identified with accessible data and a low effort and low cost analysis method.
To demonstrate so, we trace the evolution of a specific geographical region — France — from 2014 to 2021 for NDVI data in Figure \ref{figS12}. We specifically compare the images obtained for the date of 11th March to analyze the year-by-year progression.
In the images shown, a persistent increase in vegetation quality is visible in the year 2016.

\begin{thebibliography}{}
\bibitem{footnote2}{Global temperature and precipitation datasets are available through several meteorological databases (e.g. European data portal, NOAA). As a simplification, we consider higher temperatures as an indicator of general environmental stress.}
\end{thebibliography}
\end{document}